\documentclass[journal]{vgtc}

\onlineid{0}
\vgtccategory{Research}
\vgtcpapertype{Technique}

\title{Attention-Aware Visualization: Tracking and Responding to User Perception Over Time}

\author{
    \authororcid{Arvind Srinivasan$^*$}{0000-0002-3409-6077},
    \authororcid{Johannes Ellemose$^*$}{0009-0006-9676-6204},
    \authororcid{Peter W.\ S.\ Butcher}{0000-0002-3361-627X},
    \authororcid{Panagiotis D.\ Ritsos}{0000-0001-9308-3885}, 
    \authororcid{Niklas Elmqvist}{0000-0001-5805-5301}
}

\authorfooter{
\item[*] Equal contribution.
\item Arvind Srinivasan, Johannes Ellemose, and Niklas Elmqvist are with Aarhus University in Aarhus, Denmark. E-mail: \{arvind, ellemose, elm\}@cs.au.dk. 
\item Peter W.\ S.\ Butcher and Panagiotis Ritsos are with Bangor University, Bangor, United Kingdom. E-mail: \{p.butcher, p.ritsos\}@bangor.ac.uk.
}

\shortauthortitle{Attention-Aware Visualization}

\abstract{%
  We propose the notion of \textit{attention-aware visualizations} (AAVs) that track the user's perception of a visual representation over time and feed this information back to the visualization.
  Such context awareness is particularly useful for ubiquitous and immersive analytics where knowing which embedded visualizations the user is looking at can be used to make visualizations react appropriately to the user's attention: for example, by highlighting data the user has not yet seen.
  We can separate the approach into three components: (1) measuring the user's gaze on a visualization and its parts; (2) tracking the user's attention over time; and (3) reactively modifying the visual representation based on the current attention metric.
  In this paper, we present two separate implementations of AAV: a 2D data-agnostic method for web-based visualizations that can use an embodied eyetracker to capture the user's gaze, and a 3D data-aware one that uses the stencil buffer to track the visibility of each individual mark in a visualization.
  Both methods provide similar mechanisms for accumulating attention over time and changing the appearance of marks in response.
  We also present results from a qualitative evaluation studying visual feedback and triggering mechanisms for capturing and revisualizing attention.
}


\teaser{
    \centering
    \resizebox{\linewidth}{!}{\includegraphics[alt={A teaser image for Attention Aware Visualization Paper. The image is divided into three main sections: 1. Middle Section: Two figures: (a) Left figure with VR headset labeled "Data Aware 3D" with Meta Quest 3 surrounded by 3 icons. Icons represent orientation, rotation, and location. (b) Right figure with eye-tracking device labeled "Data Agnostic 2D" with Pupil Labs Neon Eyetracker surrounded by 3 icons. Icons represent gaze, pointer, and keys. 2. Left Section shows Data Aware 3D Implementation: (a) Explicit Heatmap: Shows 3D bar graphs with heatmaps. (b) Implicit Desaturation: Displays scatter plots for "Sepal Width" with gradual desaturation. (c) Always On Heatmap: Shows 3D surface plots with heatmaps for "Latitude" and "Longitude." 3. Right Section shows Data Agnostic 2D Implementation in two parts: Part 1. IllustratedPicture Framing Metaphor. Part 2. Three types of frames: (a) Bar Frame: Bar graph with maximum cumulative attention. (b) Area Frame: Area plot showing attention distribution. (c) Heat Frame: Continuous heatmap gradient.},width=\linewidth]{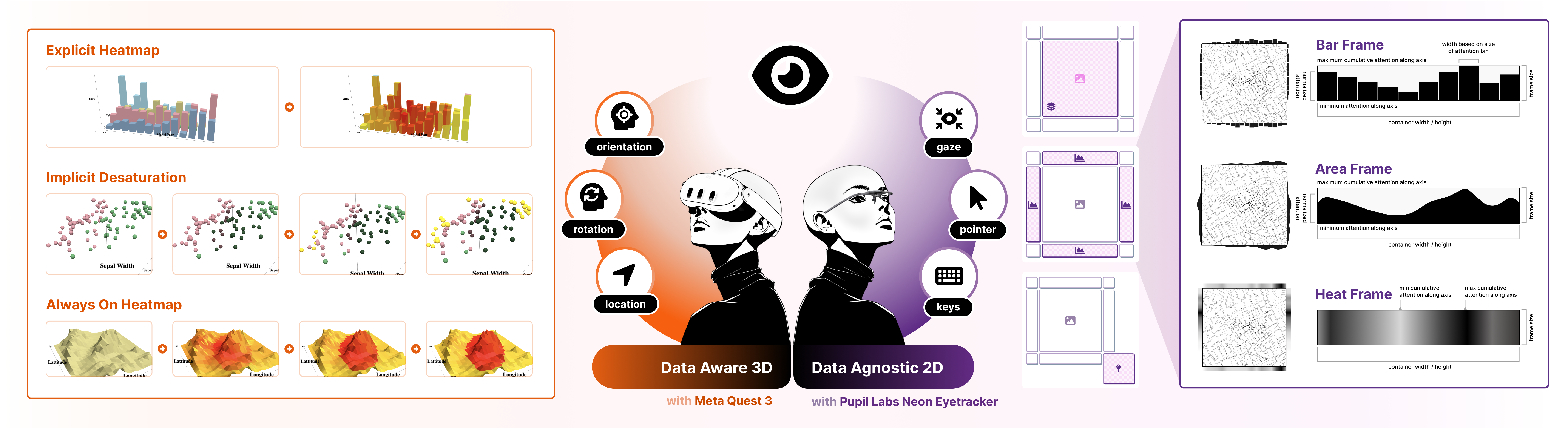}}
    \caption{\textbf{Attention-aware visualization.}
    Our work proposes tracking, recording, and revisualizing a user's attention on a visualization to aid them in understanding what they have seen and not seen, as well as cause the visualization to respond to the user's gaze.
    Our two implementations---3D and 2D, respectively---make use of the head direction of a Meta Quest 3 or similar XR HMD, the mouse pointer, or an embodied eye-tracker to measure the user's gaze on the visualization.
    We also present several methods for revisualization, including heatmaps and contour plots overlaid on the chart, or compact visualizations on the border of the viewport.
    }
    \label{fig:teaser}
}

\graphicspath{{figures/}{pictures/}{images/}{./}}

\usepackage[svgnames, dvipsnames]{xcolor}

\usepackage{enumitem}

\usepackage{svg}

\usepackage{fontawesome5}

\usepackage{booktabs}

\usepackage{listings}

\usepackage{colortbl}

\usepackage{soul}
\definecolor{new-green}{rgb}{0.104,0.667,0.229}
\newcommand{\add}[1]{{#1}} 
\newcommand{\rev}[1]{{#1}} 
\newcommand{\del}[1]{} 

\makeatletter
\newlength{\mylength}

\newsavebox{\mybox}

\makeatother

\begin{document}

\maketitle

\section{Introduction}
\label{sec:intro}

The art and science of visualization design is primarily guided by a plethora of principles and guidelines---some empirically validated, others more anecdotal---in the pursuit of crafting visual representations that are not only \textit{expressive and effective}~\cite{DBLP:journals/tog/Mackinlay86, Munzner2014} but also \textit{visually salient}~\cite{DBLP:journals/cgf/JanickeC10, matzen18visualsaliency}.
This extensive body of knowledge spans both the practical realm of visualization practitioners and the burgeoning academic discipline.
However, these visualization design principles primarily focus on creating static visual artifacts that only respond to direct user interaction, such as explicit clicks and taps.
What if we could transcend this passive paradigm to design visualizations that actively engage with the viewer's attention?
Imagine visualizations imbued with the capability to track and respond to the viewer's gaze, dynamically adapting their visual attributes in real time to optimally convey the data.

We introduce the concept of \textit{attention-aware visualizations} (AAVs): contextually aware visual representations that track user attention on their surface over time and modify their visual attributes accordingly. 
Conceptually, this approach consists of three steps (\cref{fig:teaser}): 

\begin{enumerate}
    \item\textbf{Measuring user attention} on a visualization and its marks.
    This could be as simple as estimating attention based on the scroll location or mouse cursor~\cite{Huang2012}, the user's head orientation, or their exact gaze location using an eyetracker. 
    
    \item\textbf{Recording attention over time} as the user's perception \rev{(i.e.\ gaze)} moves over the visualization \rev{over time}.
    This also includes modeling the user's short-term memory as attention smoothly decaying.
    
    \item\textbf{Modifying the visualization} in response to user attention.
    This could be used as a diagnostic---allowing the user to see which part of visualization they have surveyed and which they have not---or as a subtle visual cue for read wear~\cite{Hill1992}. 
    
\end{enumerate}

\add{While there are numerous potential approaches to designing AAVs, this paper focuses on the fundamental idea of showing users their own attention on a visualization.
Our goal is to encourage users to explore areas of a visualization they may not have previously investigated as well as to reflect on the areas they have, thus enhancing their overall understanding and interaction with the data.}

The contributions of this paper are as follows: 
(1) the concept of attention-aware visualization as a context-aware~\cite{CA_apps} method for creating visualizations responsive to implicit gaze and not just explicit interaction;
(2) a library implementing attention-awareness for 2D visualizations based on numeric integration; 
(3) an implementation of attention-awareness and occlusion~\cite{DBLP:journals/tvcg/ElmqvistT08} for 3D visualizations using the stencil buffer; and
(4) results from a \rev{qualitative evaluation} investigating the approach for both 2D (eye-tracker) and 3D settings (XR HMD).
Refer to \cref{sec:supplemental_materials} for the supplementary materials.

\section{Background}
\label{sec:background}

In this section we discuss our inspirations from prior research for this work, which include fundamental concepts such as context awareness, low-level aspects of attention in visualization, eyetracking as a measure of attention in data visualization, and the use of gaze for interaction in general HCI systems.
Note that in this paper we do not consider attention and awareness beyond the individual~\cite{DBLP:conf/cscw/GutwinG98}, such as for collaboration, territoriality, and conflict management in group work. 

\subsection{Context Awareness}

One of the primary inspirations for this work is the notion of context-aware (CA) systems~\cite{CA_apps}.
CA systems monitor and react to a user's changing context to promote and mediate said user's interaction with the system itself, as well as other systems and users. 
The term \textit{context} is defined in the literature in several ways~\cite{ALEGRE201655}, with the most popular definition from Dey~\cite[p.\ 5]{Dey_Context}, as \textit{``any information that can be used to characterize the situation of an entity,''} where \textit{``an entity can be a person, place, or object that is considered relevant to the interaction between a user and an application, including the user and applications themselves.''}
More importantly, for the context (pun intended) of visualization explored in this work, Dey~\cite[p.\ 5]{Dey_Context} defines a system as CA if \textit{``it uses context to provide relevant information and/or services to the user, where relevancy depends on the user’s task.''} 

Examples of such context range from tracking location, environmental factors, usage patterns, and user behavior.
Such context-aware systems are particularly central within the areas of mobile computing~\cite{CA_mcomp}, the internet of things~\cite{CA_iot}, augmented reality~\cite{CA_for_AR}, and wearable ubiquitous computing~\cite{AbowdWearUA, Weiser1991}.
This includes ubiquitous~\cite{Elmqvist2013} and immersive/situated analytics~\cite{Shin_TVCG_2023}, which is why we explore context awareness as a concept in both 2D and 3D in this paper.

\subsection{Attention in Visualization}

The human perceptual and cognitive systems generate a wealth of information in real time, all of which cannot be processed simultaneously.
\textit{Attention} is the selective concentration of mental awareness on a subset of this information, excluding the remaining information~\cite{Treisman1980}.
Early work on cognitive processes in attention noted that for highly complex attentive processes---such as visual search---to be possible, they must be organized in both low-level parallel---so called \textit{pre-attentive}---and complex serial---\textit{attentive}---processes~\cite{Hoffman1979, Neisser1967}.
While subsequent work has criticized the existence of a clear division between pre-attentiveness and attentiveness~\cite{Wolfe1994}, it remains a useful construct for understanding operational aspects of human perception and attention\add{, e.g., for visual search optimization \cite{lu2014subtleCues}}.

For data visualization, the work by Healey et al.~\cite{Healey1999} has been instrumental in understanding the effective use of color, shape, and motion in preattentive processing of complex visual representations.
Colin Ware~\cite{Ware2012} emphasize the cognitive limits of visual attention and memory, and provide guidelines for designing more effective visualizations that cater to human perception.
Several specialized topics related to attention has been studied within the visualization field, including change blindness~\cite{DBLP:conf/infovis/NowellHT01}, visual salience~\cite{matzen18visualsaliency, DBLP:journals/cgf/JanickeC10}, and the use of eye tracking for visualization.
We discuss the latter in more detail next.

\subsection{Eye Tracking for Visualization}

Eye tracking is a useful tool for visualization researchers~\cite{Burch2021};  it can help identify visual exploration behaviors~\cite{Burch_eyeTracking_11}, assess perceptual and cognitive aspects of engaging with visualizations~\cite{Kim_eyeTracker_12}, and even be used in combination with other usability metrics to understand user behavior~\cite{DBLP:conf/etra/BlascheckJKBE16}. 
Examples include the work of Burch et al.~\cite{Burch_eyeTracking_11} on tree-drawing perception, Alam et al.~\cite{Alam_eye_17} on eye tracking analysis in the data space instead of the commonly used image space, and Polatsek et al.~\cite{POLATSEK201826} on evaluation of salience models for visual analysis.
On the other hand, Kim et al.~\cite{Kim_eyeTracker_12} assess the limitations of eye trackers in information visualization studies and find that in certain situations inferences about the underlying cognitive processes may be misleading.
Kurzhals et al.~\cite{Kurzhals_16} present a survey of eye tracking technologies for the evaluation of visualization techniques, identifying future directions at the time of publication.
In more recent work Shin et al.~\cite{ShinScanner23} review prior work on eye tracking and propose a novel approach of creating a virtual eye tracker that produces gaze heatmaps using deep learning. 

Another take on visualization and eye tracking is the use of visualization for eye tracking data.
Blascheck et al.\ provide a somewhat dated survey on this practice from 2014~\cite{DBLP:conf/vissym/BlascheckKRBWE14}.
Notable examples include Tsang et al.'s approach for visualizing fixation sequences~\cite{DBLP:journals/tvcg/TsangTS10},
Alam and Jianu~\cite{DBLP:journals/tvcg/AlamJ17} use visualization to connect gaze patterns with visual elements on a screen, and the VETA system~\cite{DBLP:journals/vi/GoodwinPWHLAD22} combines eye tracking visualization with video of the underlying scene.
Our work in this paper both uses eye tracking (in addition to other user input, such as touch events and mouse actions) to collect the context from the user as well as visualize attention collected from eye tracking data on a visualization.
Our attention-aware visualization approach both builds on this past work and extends it significantly by using the eye tracking data as a first-class input channel and not merely a diagnostic tool; more below.

\subsection{Gaze as Input}

The use of gaze as a method for computer input is marked by pioneering efforts to establish its viability and utility.
Bolt~\cite{DBLP:conf/siggraph/Bolt81} demonstrated one of the earliest implementations, showing how dynamic window systems could be orchestrated through gaze.
This foundational work led to followup efforts, including Ware and Mikaelian~\cite{DBLP:conf/chi/WareM87}, who evaluated eye tracking for computer input, and Jacob's proposal of eye movement-based techniques that link gaze direction with interface control~\cite{DBLP:conf/chi/Jacob90}.

As gaze tracking technology matured, researchers began exploring its integration with other input modalities to create richer, more nuanced interaction paradigms.
Zhai et al.'s MAGIC pointing~\cite{DBLP:conf/chi/ZhaiMI99}, Salvucci and Anderson's gaze-added interfaces~\cite{DBLP:conf/chi/SalvucciA00}, and Fono and Vertegaal's EyeWindows~\cite{DBLP:conf/chi/FonoV05} each contribute to this domain by cascading gaze with manual inputs to streamline selection tasks and manage focus within interfaces.
Pfeuffer et al.\ introduced gaze-touch in 2014~\cite{DBLP:conf/uist/PfeufferACG14}, combining gaze with multi-touch on the same surface and thereby illustrating the synergistic potential of gaze with other interaction techniques.

\rev{Over the past decade, there has been significant progress in gaze-supported multimodal interactions, particularly for exploring large datasets and improving target acquisition.}
Stellmach et al.~\cite{DBLP:conf/ngca/StellmachSND11} and Stellmach and Dachselt~\cite{DBLP:conf/chi/StellmachD12} explored gaze-supported interactions for large image collection exploration and target acquisition, respectively. Turner et al.~\cite{DBLP:conf/interact/TurnerABSG13}, Pfeuffer et al.~\cite{DBLP:conf/uist/PfeufferACG14, DBLP:conf/uist/PfeufferG16}, and Velloso et al.~\cite{DBLP:conf/interact/VellosoTABG15} further investigated the integration of gaze with touch and mid-air gestures, demonstrating the potential for seamless and efficient interactions across devices and interfaces. Pfeuffer et al.’s work on combining gaze with pinch interactions in virtual reality~\cite{DBLP:conf/sui/PfeufferMMG17} represents the latest evolution of this research direction, highlighting the growing sophistication and breadth of gaze-supported interactions.\footnote{Incidentally, gaze and pinch has also been adopted by the Apple Vision Pro, making it the first commercial device using dedicated gaze interaction.}
Although this research is closely related to AAVs, our approach focuses less on utilizing direct gaze as an input to control an interface, \add{as explored in \cite{Lu2020glanceableAR}}, and more on the retrospective analysis of gaze patterns over time.

\section{Design: Attention-Aware Visualization}
\label{sec:aav}

The intellectual contribution proposed in this work is the concept of \textit{attention-aware visualizations} (AAV): data visualizations that track the viewer's attention on their surface over time and react accordingly. 
\rev{We see several benefits to this approach, ranging from creating a more responsive and engaging visual environment, to showing data coverage, illuminating missed outliers, and even predicting attention patterns.}
\add{Although numerous possibilities exist for how a visualization could respond to a viewer’s attention, this paper focuses solely on displaying the viewer’s attention on the visualization.}
We can summarize our approach using the following research questions:

\begin{itemize}
    \item[RQ1] How should a viewer's attention on a visualization be \textbf{measured in real time}? 
    \item[RQ2] How should a viewer's attention on a visualization be \textbf{recorded over time}? 
    \item[RQ3] How should a visualization be \textbf{modified to display} the viewer's attention over time? 
\end{itemize}

Here we propose our design framework for attention-aware visualizations that addresses each of these research questions in turn.

\begin{figure}[htb]
    \centering
    \includegraphics[alt={}, width=\linewidth]{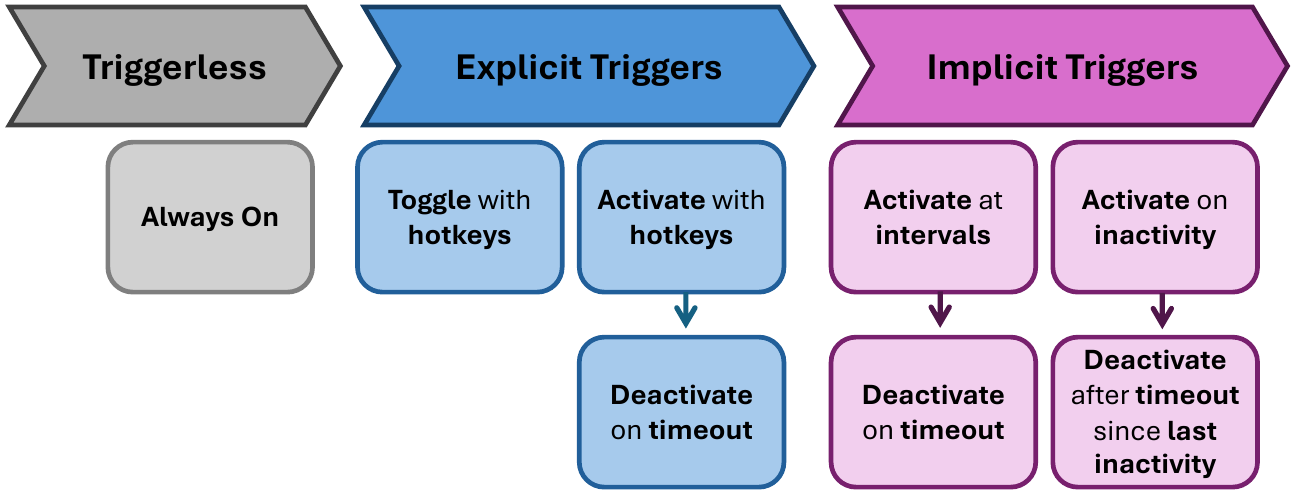}
    \caption{\textbf{Triggering strategies.}
    Overview of the different triggering strategies for our attention-aware visualization interventions.}
    \label{fig:triggering}
\end{figure}

\subsection{Measuring Attention}

First of all, what is \textit{attention}?
As discussed in the Background (\cref{sec:background}), there are several interpretations. 
From a cognitive perspective, attention refers to the selective focusing of mental resources on specific information or stimuli.
Even though this view encompasses all senses, we focus on vision here since most interactive visualizations primarily leverage the visual system. 
Attention, then, determines which aspects of the visual scene are prioritized for mental processing, guiding where and how we allocate our visual resources.

Given all that, how should we measure it (RQ1)?
Since it is a purely cognitive process, attention cannot be measured in its own right---at least not with widely available equipment.
However, depending on the context when using a computing device, there are several situations when interaction with the device can be used to measure attention:

\begin{itemize}
    \item[\faHandPointer]\textit{Taps on a touch screen} signifies an interest in that area (if the touch was intentional);
    \item[\faMousePointer]\textit{Pointer movement} on the area surrounding the pointer or its destination (if the pointer is being used deliberately); and
    \item[\faEye]\textit{Eye movement} on the screen (if the user is viewing the device actively and not preoccupied by some other task).
\end{itemize}

We have experimented with all forms of attention measurement. 
They all have their particular uses, especially since some of them are less available than others. 
For example, eye movement is indisputably the most useful metric, but capturing user gaze requires a dedicated eyetracker to achieve high accuracy, and these are typically expensive and not widely available. 
Mouse movement, on the other hand, is trivially available on most computing platforms, but may not always be deliberate---and it can be hard to determine when this is the case.

\subsection{Recording Attention}

We assume a reliable attention measurement mechanism that reports the user's attention on a visualization in real time.
We further assume that the attention focus is a specific point in the user's point of vision, presumably with a specific area surrounding it and with some degree of uncertainty depending on the measurement device (e.g., an embodied eyetracker being much more accurate than a webcam-based eyetracker).

Now, how do we record this attention over time (RQ2)?
At this point, we can identify two separate approaches to recording attention depending on whether we have knowledge of the visualization or not:

\begin{itemize}
    \item\textbf{Data-agnostic:} We do not have specific knowledge of the underlying data in the visualization being tracked---in other words, we do not know what the pixels of the visualization represent, only that the user's attention is moving over it; and
    
    \item\textbf{Data-aware:} We have specific knowledge of exactly what parts---what marks representing specific data items---of the visualization that the user is focusing their attention on.
\end{itemize}

\paragraph{Data-agnostic.}

Since we have no knowledge of the underlying visualization or its data, we cannot directly associate attention with a specific mark or data item. 
Rather, this approach requires subdividing the visual space of the visualization into discrete bins---typically square cells on a regular grid---that we can use to accumulate attention over time.
Unfortunately, this limits our chances of detecting when the user's attention is drifting because we cannot detect situations when they are fixating on empty space in a visualization.

\paragraph{Data-aware.}

Because we can associate points on the visualization with visual marks, we can directly accumulate user attention on these marks (or even parts of marks).
We can also discard attention that is spent on empty parts of a visualization---although we naturally still cannot reliably understand when the user is absent-mindedly viewing the visualization without actually seeing it.

\paragraph{Representing attention.}

Each \textit{attention target}---a cell in the agnostic case or a visual mark in the aware case---can be assigned an attention value that starts at zero and then accumulates for every time unit the user's area of attention overlaps with the cell or mark.
If more than one attention metric is active at the same time---say, pointer and gaze---then each target has two attention values.
Finally, attention should be normalized or capped across the entire visualization so that viewing one target for a long time will not overshadow all other targets.

\paragraph{Attention decay.}

People have limited short-memories and thus attention decays over time.
In other words, just because a person has spent a long time looking at a specific set of bars in a barchart does not mean that they will be indelibly fixed in the person's mind. 
After a while looking at other parts of the chart, high attention values for specific attention targets should decay towards zero over time.

\subsection{Revisualizing Attention}

How do we actually use the accumulated attention metric to change the visualization (RQ3)?
One option is to not change the visualization at all, but to merely use the data for evaluation purposes where the viewers themselves do not ever see it.
However, in this paper, we are interested in investigating how this data can be utilized by the visualization itself, \add{e.g. to encourage exploration, or facilitate introspection}.
We call this \textit{revisualizing} the attention on the original visualization.

\paragraph{Presentation.}

There are several potential methods of presenting the accumulated attention metric. 
They differ in how invasive they are to the target visualization:

\begin{itemize}
    \item[\faCircle] \textbf{Marks:} Modify the original marks to convey attention; 
    \item[\faLayerGroup] \textbf{Overlay:} Visualize attention on an overlay on top of the view;
    \item[\faChartArea] \textbf{Border:} Use view border\rev{s} to visualize attention; and 
    \item[\faMapPin] \textbf{Minimap:} Visualize attention on a separate window.
\end{itemize}

There are many options for the visual representation for the attention metric. 
Their availability depends vastly on whether the approach is data-agnostic or data-aware.
For the former, the marks themselves cannot be changed since we have no knowledge of the underlying visualization.
Instead, we have to use image processing operations on cells to convey the data, such as color transforms (hue rotate, desaturation, and luminance), blur, transparency, heatmaps, etc.
For the latter, we can modify the visual attributes of the marks using object-level operations.

\paragraph{Triggering.}
\label{sec:triggering}

Finally, because visualizing attention while tracking attention easily becomes a circular and self-reinforcing system, we must be careful to consider when and how to trigger these techniques.
\Cref{fig:triggering} gives an overview of our proposed triggering strategies.
A fully na{\"i}ve solution would be triggerless and ``always-on,'' both for the attention collection as well as the revisualization.
However, this would invariably mean that as the revisualization updates in response to the user's attention, the change will draw the user's attention, and so on.
This will lead to an unstable and potentially confusing display.

We further delineate two additional strategies: \textit{explicit} vs.\ \textit{implicit} triggers.
The former would require an explicit action to activate on behalf of the user; for example, checking a box or pressing a hotkey.
This has the disadvantage that the trigger is outside the main workflow of the visualization and may thus be forgotten and never triggered.
The latter approach would activate not by user action, but by some external implicit event, such as user inactivity, regular intervals, or perhaps the presence of consistently overlooked parts of the data.
It makes sense that for both these triggers, attention data collection is disabled while the visualization is active to prevent reinforcing loops.

\section{Implementing Attention-Aware Visualization}
\label{sec:impl}

We have implemented two versions of the attention-aware visualization method (\cref{sec:aav}), one for 2D settings using the data-agnostic approach, the other for 3D and the data-aware approach.
Providing two separate implementations enables us to explore two quadrants in the design space in \cref{tab:aav-impl}.
The other two quadrants in the design space have not been implemented, but the extension should be trivial.

At the core of this endeavor lies a common model for measuring, recording, and revisualizing attention for both implementations.
In this section, we describe this common model.
In the following two sections, we describe the 2D and 3D implementations in detail, respectively.

\colorlet{RowColor}{SteelBlue!20}

\begin{table}[htb]
    \centering
    \renewcommand{\arraystretch}{1.2}
     \sffamily 
     \small
    \caption{\textbf{AAV implementations.}
        We provide two AAV implementations; a data-agnostic one for 2D, and a data-aware one for 3D.
    }
    \label{tab:aav-impl}
    \begin{tabular}{c|cc}
        \hline
        \rowcolor{RowColor}
        \cellcolor{SteelBlue!50} & \textbf{Data-agnostic} & \textbf{Data-aware}\\
        \hline
        \cellcolor{RowColor}\bf 2D & AAV2D (\cref{sec:2d-impl}) & -- \\
        \cellcolor{RowColor}\bf 3D & -- & AAV3D (\cref{sec:3d-impl})\\
        \hline
    \end{tabular}
\end{table}

\subsection{Attention Model}
\label{sec:attention-impl}

Drawing on~\cref{sec:aav}, the common attention model underpinning both of our implementations conceptualizes attention as a transient, ephemeral attribute that mirrors the characteristics of human short-term memory.
This model is based on measuring the user's gaze at specific elements within the visualization---whether these are discrete \textit{bins} in a data-agnostic approach, or distinct visual \textit{marks} (representing data points) in a data-aware framework.
The attention allocated to these elements will accordingly escalate at a specific rate.
The increment reflects the elements' entry into the viewer's short-term memory, akin to how prolonged observation may enhance memorability.
However, once the viewer's gaze shifts away, the model dictates that attention diminishes at a predetermined decay rate, analogous to the fading of details from short-term memory.

In practice, our model manages two \textit{attention maps} for the visualization:
(1) the global cumulative attention, and (2) the current short-term attention. 
For a data-aware implementation, an attention map is an associative array recording attention value per data point.
For a data-agnostic one, an attention map is a spatial data structure (2D planar or 3D volumetric grid) recording attention per spatial cell in the visualization substrate.
The global cumulative attention map does not decay over time, but keeps track of the attention measure over the entire lifecycle of the visualization.
It is used primarily as a diagnostic and is not directly revisualized to the user (except potentially at the end of a session).
We use the cumulative attention map to update the short-term attention map; these current attention values will ebb and flow as the user's gaze moves across different parts of the visualization.
This short-term attention map is used to determine both \textbf{when} (triggering) and \textbf{how} (revisualization) to render the user's attention on the visualization.

\subsection{Triggering Mechanisms}
\label{sec:impl-triggering}

As noted in~\cref{sec:triggering}, we identify three strategies for triggering the revisualization of the visualization: 
\textsc{Always-on}, \textsc{Explicit} triggering, and \textsc{Implicit} triggering (\cref{fig:triggering}). 
\textsc{Always-on} is constantly active.
Depending on the revisualization approach, this behavior may lead to the revisualization itself affecting the user's attention. 

For \textsc{explicit} triggering, we activate revisualization when the user presses a hotkey or controller button---potentially spring-loaded, so that when the button or key is released the revisualization disappears.
Furthermore, as long as the revisualization is visible, we temporarily disable capturing attention to avoid self-reinforcing behavior.

Finally, for \textsc{implicit} triggering, we distinguish between two forms of revisualization: \textit{emphasis} and \textit{de-emphasis}.
When the short-term attention for an entity (bin or mark) has decayed below a lower threshold, this means that there is a risk that the user is not seeing the entity, so the emphasis revisualization is activated.
Analogously, when the short-term attention has increased above an upper threshold, the de-emphasis revisualization will be activated to discourage further study. 
Once the value rises above the lower or dips below the upper thresholds, the revisualization is de-activated.
Specific implementations may choose to disable the emphasis or de-emphasis revisualizations as needed.

\subsection{Revisualizing Attention}
\label{sec:revis-impl}

Finally, we use a common revisualization approach for both our 2D and 3D implementations.
Common between them is the use of a straightforward heatmap with a quantitative color scale to revisualize attention. 
Furthermore, as discussed above, the notion of emphasis and de-emphasis is used for implicit triggering, and is similar across both implementations. 
We use \textit{color saturation} as a common denominator for both: reducing saturation to de-emphasize and increasing it to emphasize.
In addition, as discussed below, in the 2D case we also support blur to de-emphasize a segment of a visualization.

In the 3D implementation, we apply all of our revisualizations on the 3D scene itself; see~\cref{sec:3d-impl} for detail.
However, in the case of the 2D implementation, the visualizations are not immersive but bounded in 2D space, which allows us to revisualize attention also outside the visualization itself. 
This has the added benefit of allowing us to disable capturing attention when the user's gaze is directed outside of the visualization itself.
See~\cref{sec:2d-impl} for more details.

\add{The use of color to encode the revisualization of course means that color cannot be used to simultaneously encode dimensions of the data, which is not desired. 
However for immersive environments, using e.g. scale to denote attention would skew the perception of the data.
For this paper we consider it a necessary compromise, and using an \textsc{explicit} triggering approach still allows for encoding data using color.}

\begin{figure*}[htb]
    \centering
    \begin{minipage}{0.99\columnwidth}
        \centering
        \subfloat[Glaze overlay.]{
            \includegraphics[width=0.32\linewidth]{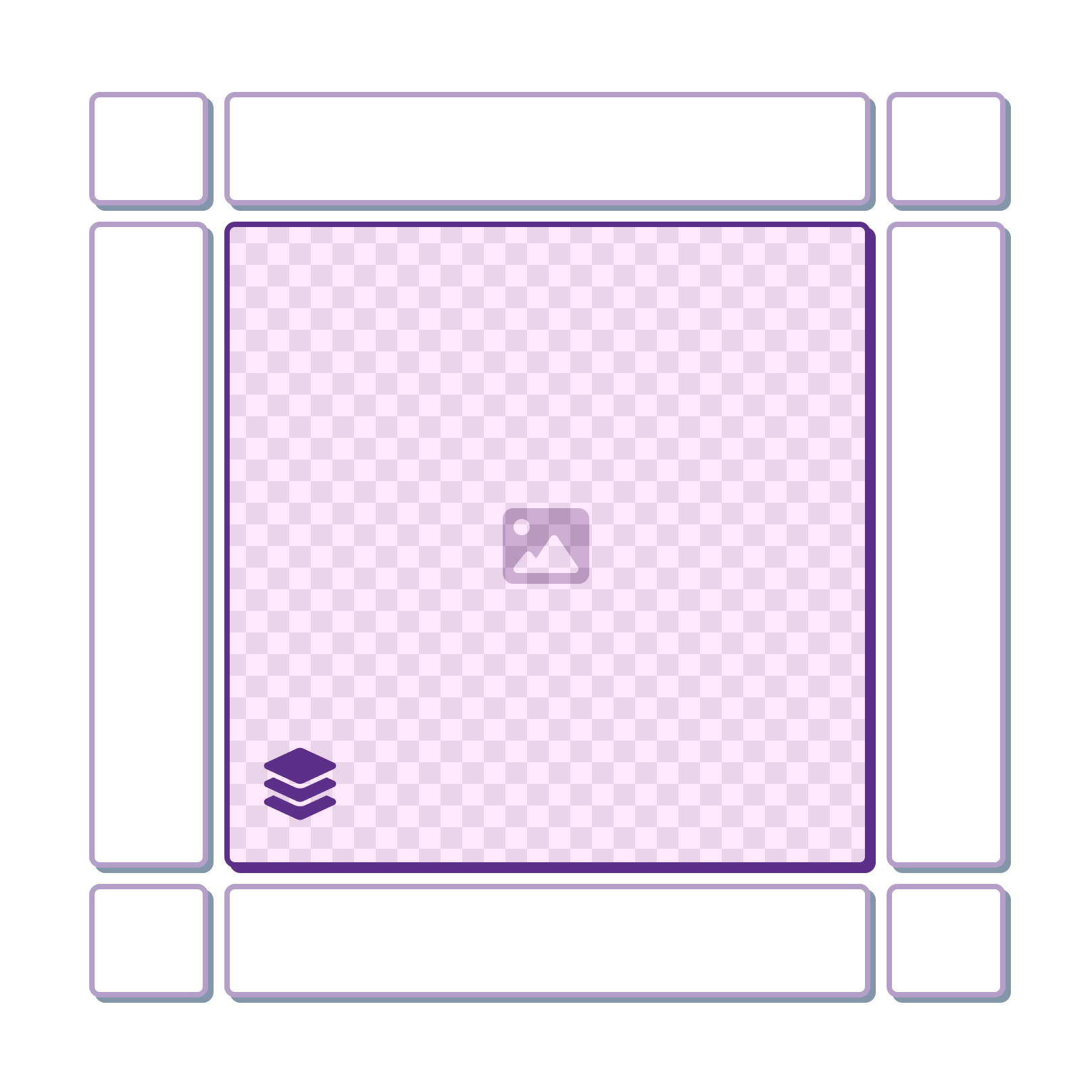}
            }
        \subfloat[Minimap on the border.]{
            \includegraphics[width=0.32\linewidth]{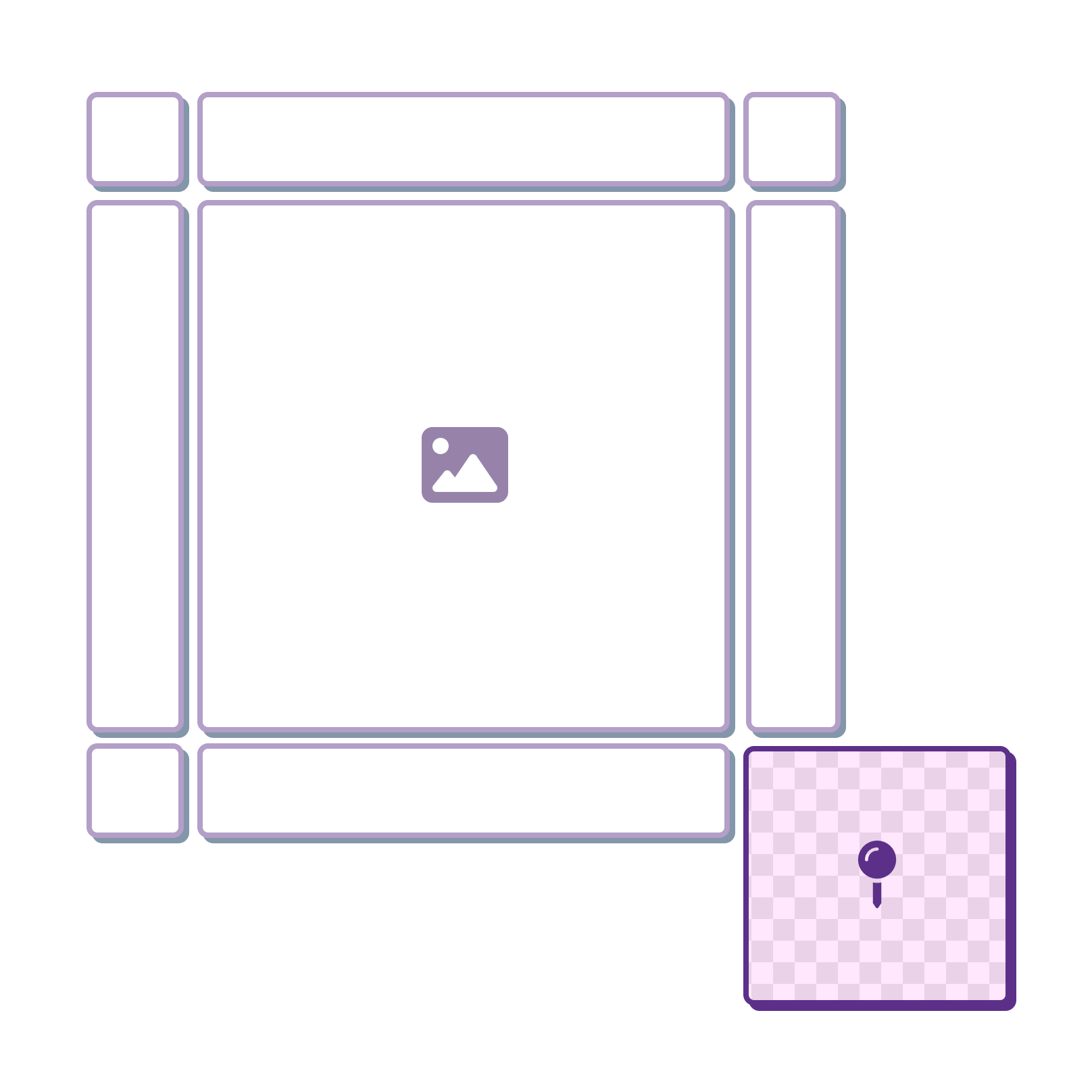}
            }
        \subfloat[Decorative frame border.]{
            \includegraphics[width=0.32\linewidth]{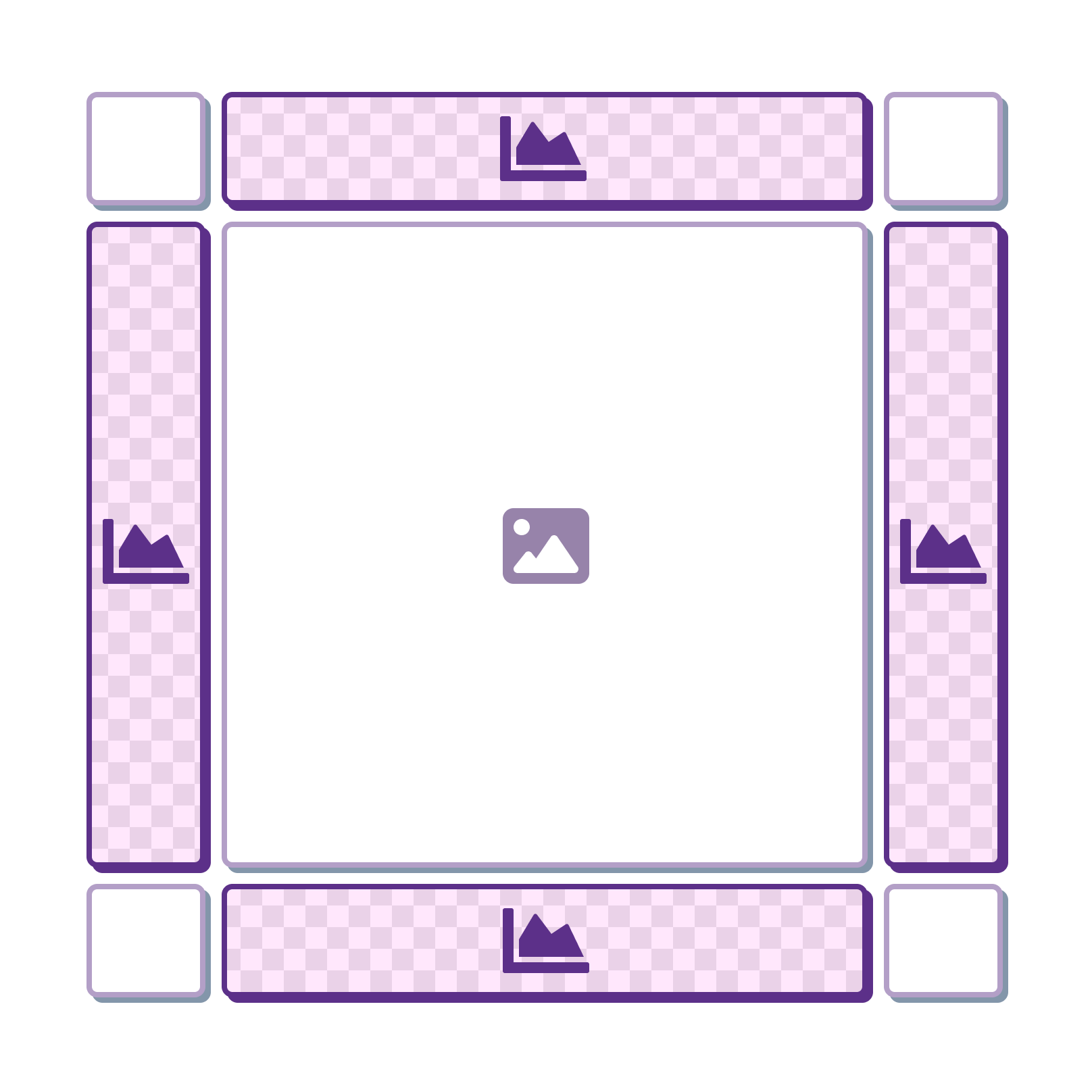}
            }
        \caption{\textbf{AAV2D picture framing metaphor.}
        AAV2D provides several methods for revisualizing attention as an (a) {\color[HTML]{5C3089}\faLayerGroup\textsc{ \textbf{Overlay}}}, either on the {\color[HTML]{5C3089}\faImage\textsc{ \textbf{Mount}}} that holds the visualization or on a (b) {\color[HTML]{5C3089}\faMapPin\textsc{ \textbf{Minimap}}} around the (c) decorated {\color[HTML]{5C3089}\faChartArea\textsc{ \textbf{Border}}} that revisualizes attention.}
        \label{fig:aav2d-framing}
    \end{minipage}
    \hspace{0.02\textwidth}
    \begin{minipage}{0.99\columnwidth}
        \centering
        \subfloat[Heatmap.]{
            \includegraphics[width=0.32\linewidth]{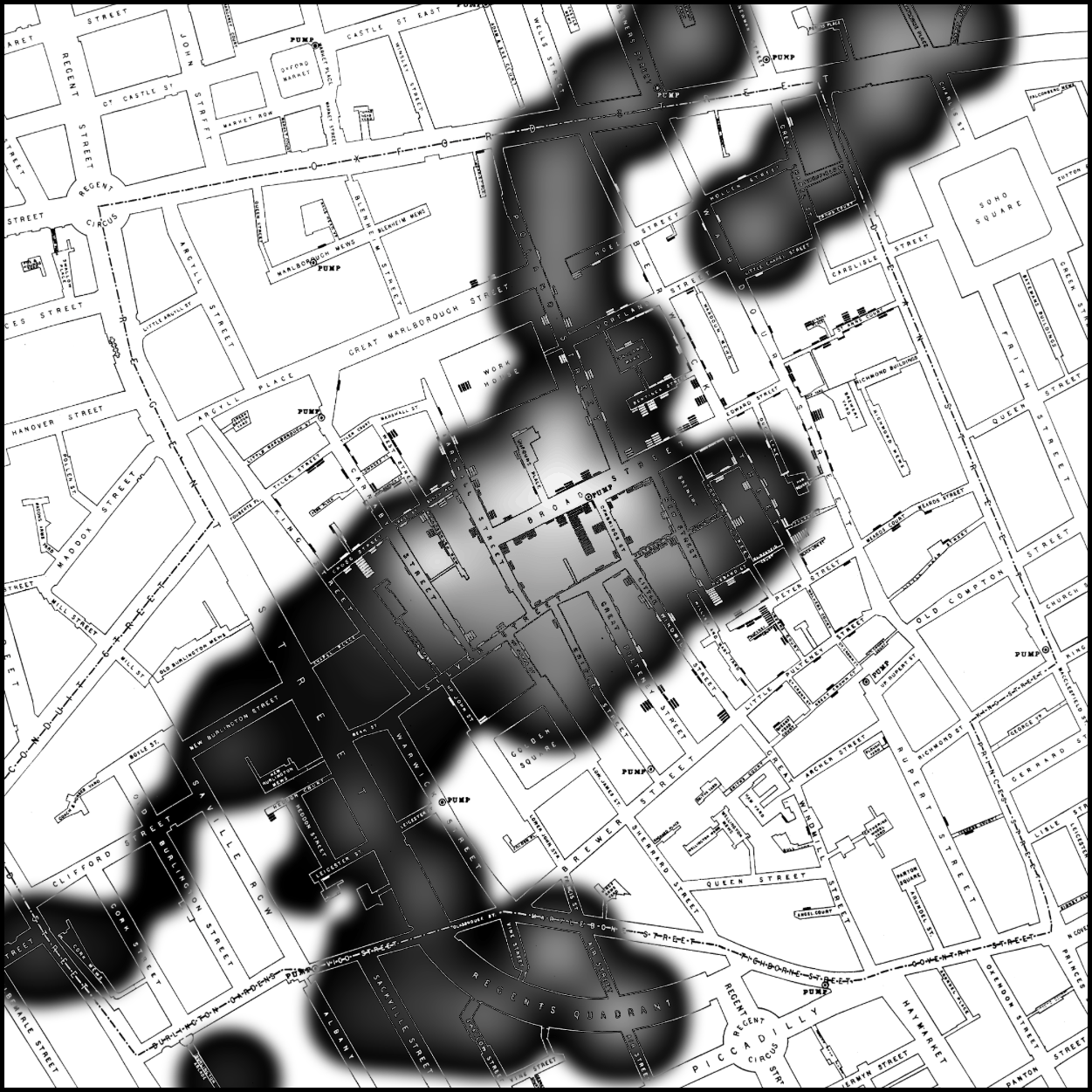}}
        \subfloat[Contour plot.]{
            \includegraphics[width=0.32\linewidth]{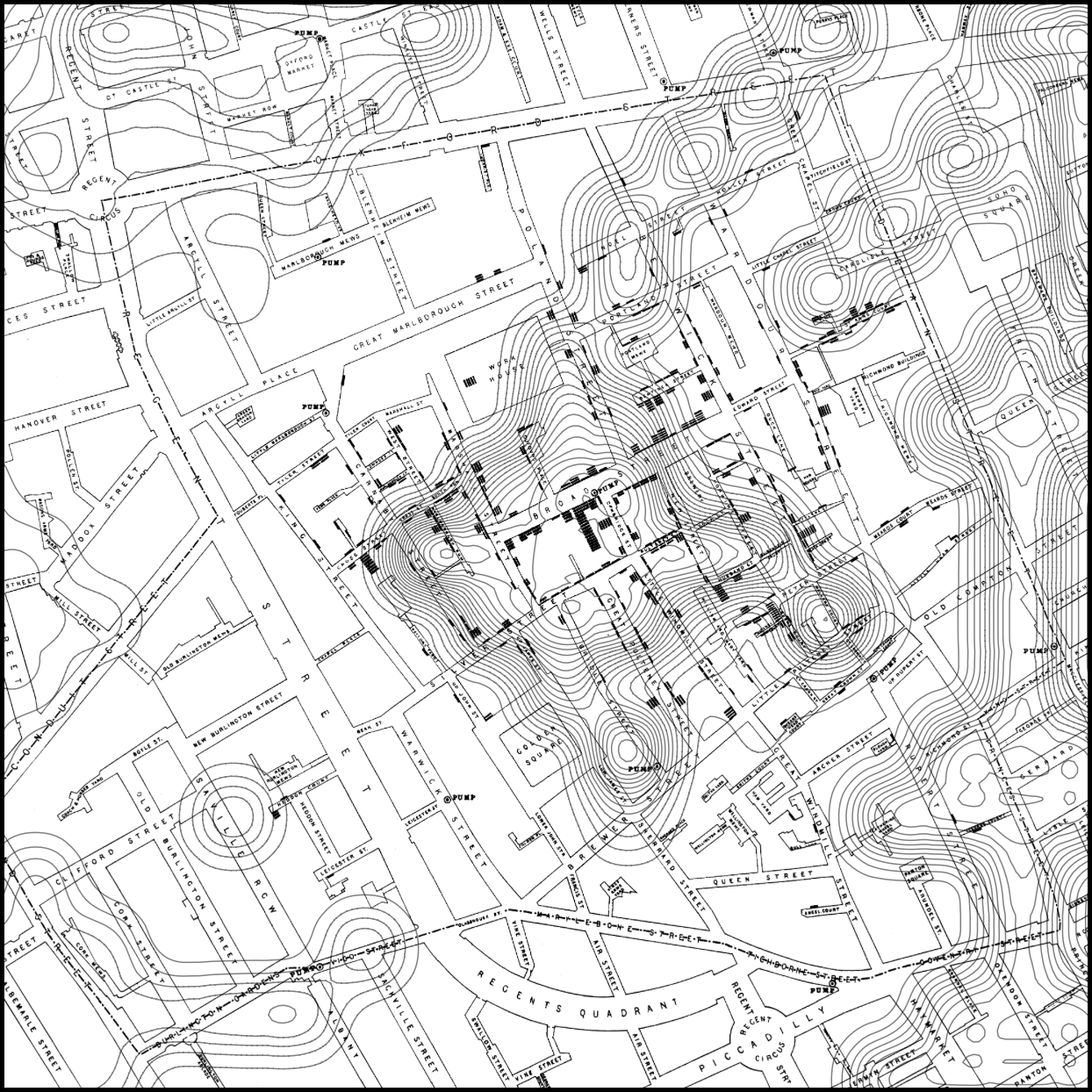}}
        \subfloat[Mesh.]{
            \includegraphics[width=0.32\linewidth]{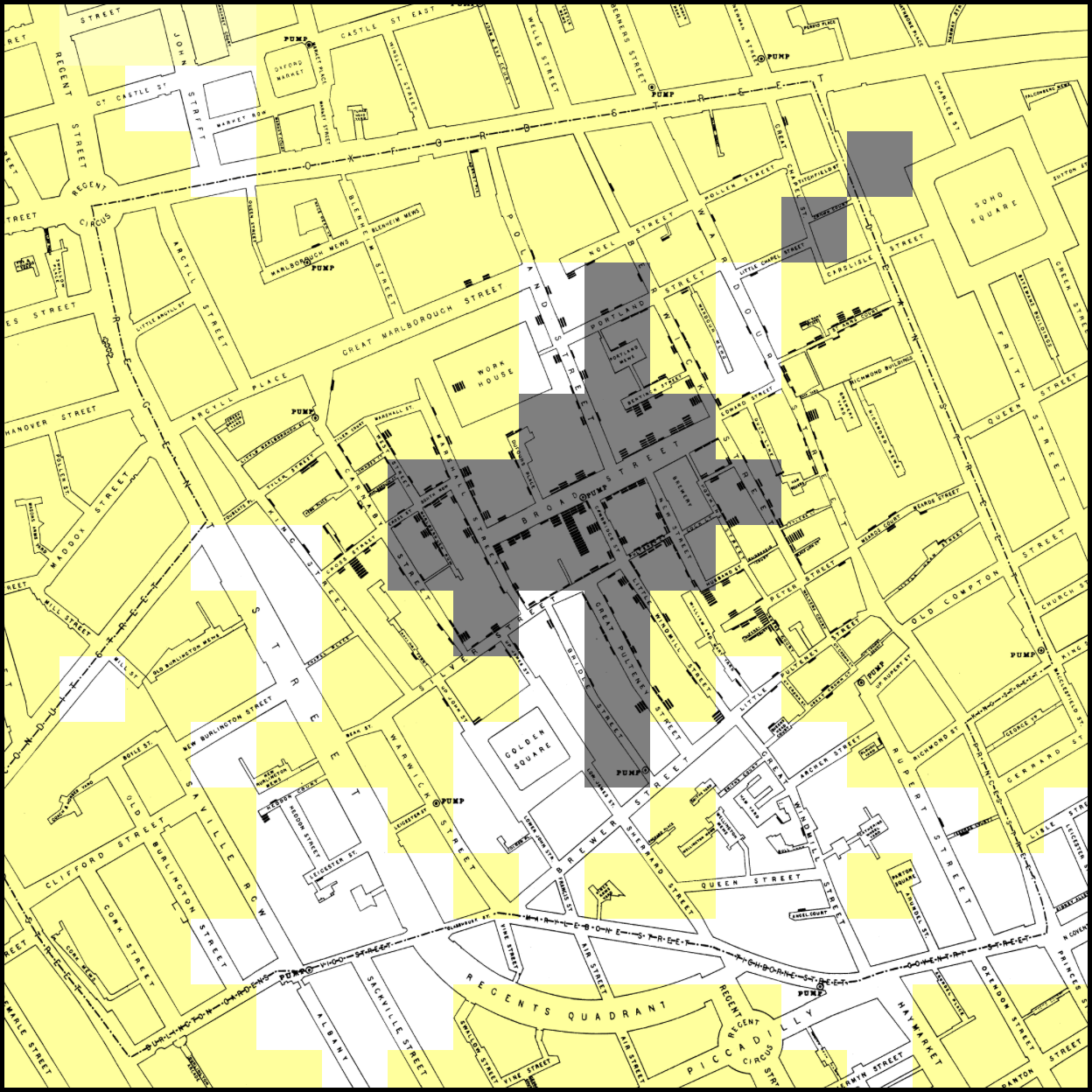}}
        \caption{\textbf{AAV2D overlay (glaze) revisualizations.}
        The three primary revisualization mechanisms include (a) \textbf{heatmap}, (b) \textbf{contour plot} and (c) \textbf{mesh}.
        The colors shown in the figure are tailored to fit the corresponding mounted visualization, which in this case is \textit{John Snow's Cholera Map}.
        }
        \label{fig:aav2d-revis}
    \end{minipage}
\end{figure*}
\section{Data-Agnostic AAV in 2D}
\label{sec:2d-impl}

AAV2D is our 2D implementation of attention-aware visualization.
It is a data-agnostic JavaScript package made available on \texttt{npm}.
Because it is data-agnostic, it can decorate any given HTML element in a DOM and decorate it with attention capture and visualization.
Below we discuss attention capture, revisualization methods, and implementation.

\subsection{Attention Capture}

The AAV2D package maintains both a global cumulative and a short-term current attention map, as described in \cref{sec:attention-impl}, in the form of a 2D grid fitted to the attached element.
The size of individual grid cells can be controlled through programming-level and user-level options; see below.
The AAV2D package installs input event handlers on the element, including both mouse and eye-tracker events (if available). 
This makes it possible to track the user's gaze or mouse pointer position whenever it moves within the bounds of the element. 
For both approaches, the user's attention point is treated like the center of an ``attention circle,'' and the package accumulates attention to all cells intersected by this circle.
Attention data outside the region is ignored. This 2D grid emits events at the individual cell level, enabling the users of this library to create custom renderers if they so desire, similar to ones used in the study with the corresponding triggering mechanisms.

\subsection{Revisualization}

Our AAV2D implementation uses a ``picture framing'' metaphor for its revisualization mechanism, supporting several of the presentation methods in~\cref{sec:revis-impl}.
More specifically, attaching AAV2D to an HTML element is seen as \textit{mounting} a picture.
The Overlay presentation is a \textit{glaze} applied on top of the picture.
The library also provides a Border presentation that serves as the \textit{decorative frame} of the picture.
Finally, the \textit{Minimap} is provided as another decoration on the frame.
The frame is also used for showing an options panel that can be toggled to be visible or not.
\Cref{fig:aav2d-framing} shows a visual overview of this method.

\begin{figure}[!htb]
    \centering
    \subfloat[Bar chart.]{
        {\includegraphics[width=\linewidth]{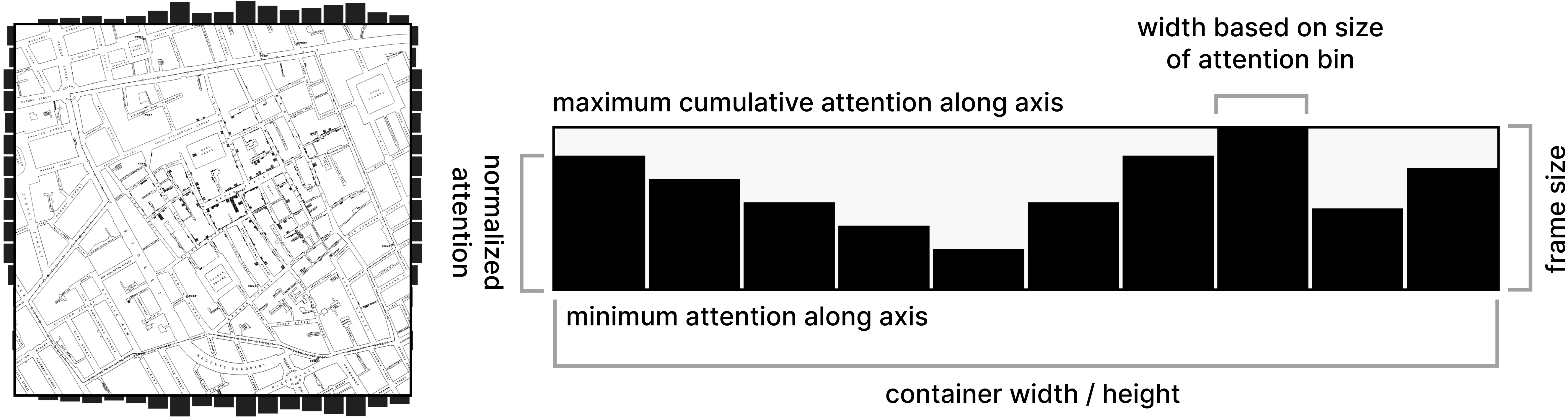}}}\\
    \subfloat[Area chart.]{
        {\includegraphics[width=\linewidth]{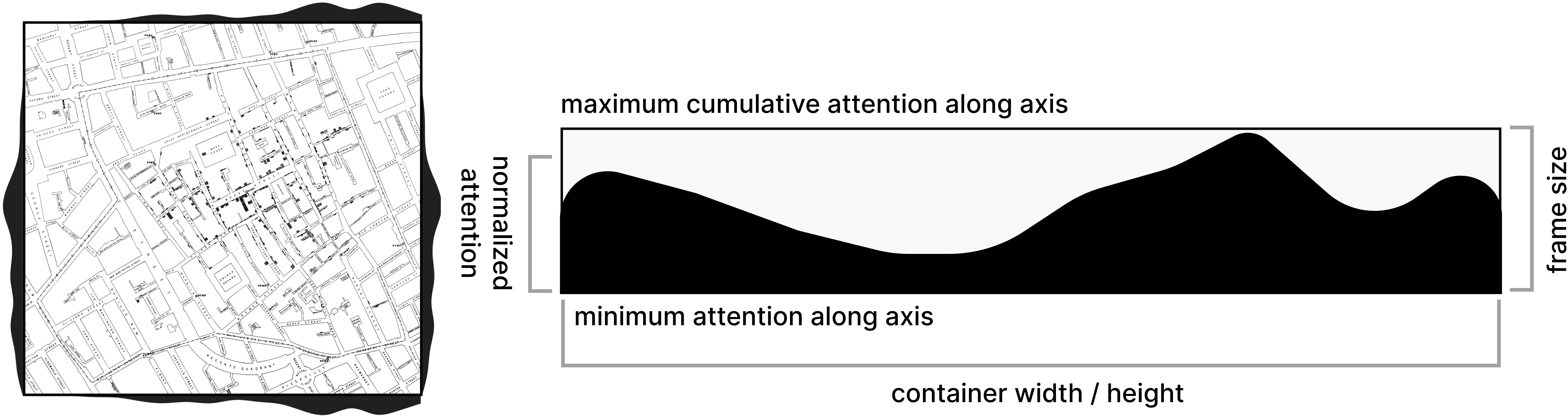}}}\\
    \subfloat[Linear heatmap.]{
        {\includegraphics[width=\linewidth]{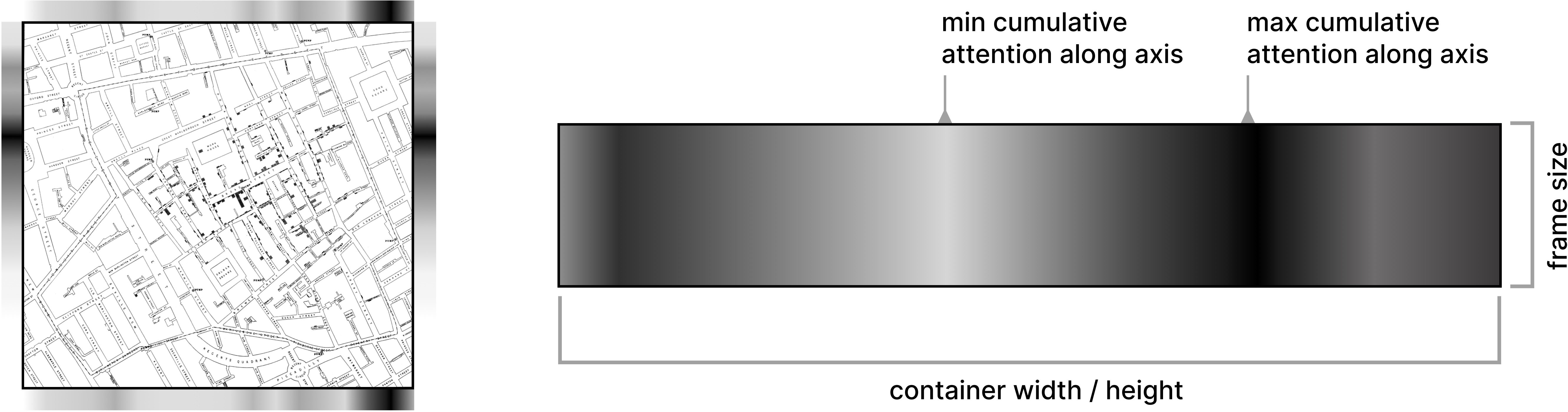}}}
    \caption{\textbf{AAV2D border revisualizations.}
    AAV2D provides several methods for revisualizing attention on the decorative border of the element, each of which can be tailored to fit the mounted visualization.
    }
    \label{fig:aav2d-border}
\end{figure}

Furthermore, AAV2D supports several revisualization approaches depending on the presentation method.
For both the glaze overlay and border minimap, the approach can use a heatmap, contour plot, or \add{mesh} (\cref{fig:aav2d-revis}); in the case of the overlay, this revisualization is shown on top of the element, whereas for the minimap, it is shown on the miniature.
For the frame border, we support bar charts, area charts, and heatmaps that show the short-term or cumulative attention on the plot for a specific $x$ or $y$ coordinate; see \cref{fig:aav2d-border} for examples.

\subsection{Implementation Notes}

We implemented AAV2D using vanilla JavaScript without the use of specialized libraries or bindings.
D3\footnote{\url{https://d3js.org}}~\cite{D3} was used for composing Scalable Vector Graphics (SVG) of contour and heatmaps.
The mesh implementation uses CSS image filters to perform some of its visual transformations on the underlying HTML element, such as blur, saturation, and hue-rotation.
All input events are intercepted by AAV2D event handlers, but are then propagated to the underlying element, meaning that it can be used even for interactive visualizations and interfaces.

\begin{figure}[!htb]
    \centering
    \subfloat[Explicit Heatmap.]{
        \resizebox{\linewidth}{!}{\includegraphics[width=\linewidth]{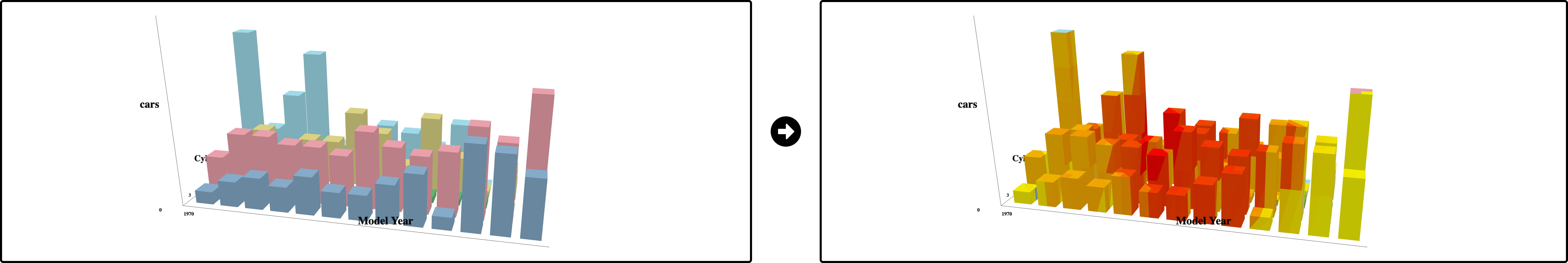}}
    }\\
    \subfloat[Implicit Desaturation.]{
        {\includegraphics[width=\linewidth]{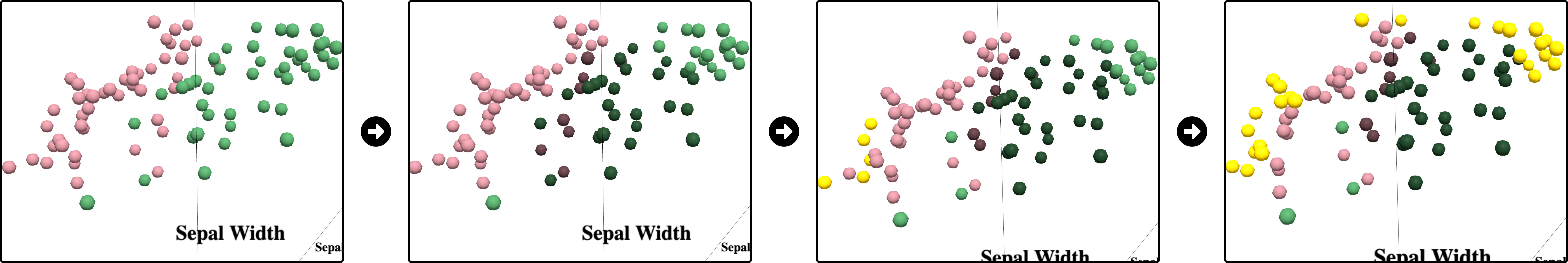}}      
    }\\
    \subfloat[Always On Heatmap.]{
        {\includegraphics[width=\linewidth]{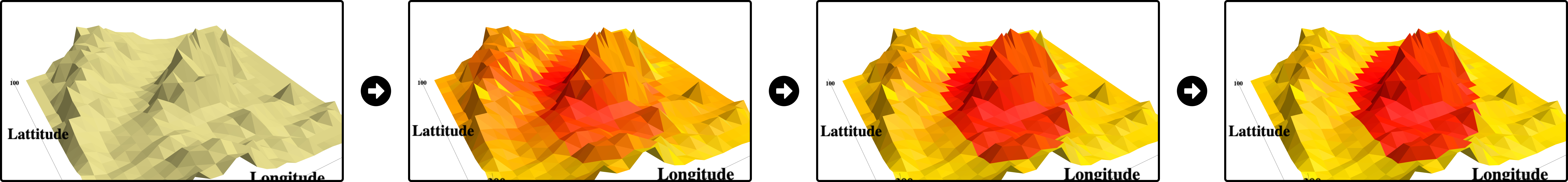}}      
    }
    \caption{\textbf{AAV3D revisualizations.}
    AAV3D provides several methods for revisualizing attention on the visual marks themselves.
    }
    \label{fig:aav3d-impl}
\end{figure}

\section{Data-Aware AAV in 3D}
\label{sec:3d-impl}

Here we present AAV3D, our 3D data-aware implementation of attention-aware visualizations suitable for immersive analytics settings.
Below we discuss the attention capture and triggering, revisualization, and notes about our implementation.

\subsection{Attention Capture}

The implementation has two strategies for capturing attention: using screen coordinates, and using the center of the screen. 
The screen coordinates strategy allows attention to be captured for elements currently rendered at any coordinate of the screen. 
This is useful for a non-immersive setting, where the interactive 3D scene is rendered on a normal display. 
In such a setting, the screen coordinates can be found with an eye-tracker, or by using the cursor as a proxy for gaze. 

The second strategy only apply attention to elements in the center of the screen.
This strategy is specifically designed for immersive environments using XR headsets that do not include eyetracking, such as the Meta Quest 3\add{ (which was the target implementation platform for AAV3D)}. 
In such immersive environments, the user's head orientation thus acts as a proxy for their attention. 

Common for both of these strategies is that they apply attention to all elements that are within a specified distance from the user's gaze position or center of the screen. 
This allows attention to be captured for the general area that the user's attention is directed. 
By carefully selecting an appropriate distance, we can approximate the user's attention from their head orientation in XR. 

Beyond recording attention to individual visual marks, we can capture attention at the level of individual 3D triangles based on their visibility to the user.
This can be useful to deal with occlusion arising from other objects in the scene, or from hidden surfaces facing away from the user.
We go into details on how this is done in \cref{sec:3d-impl-notes}.

\subsection{Revisualization}

Our implementation currently supports two types of revisualizations:

\begin{itemize}
    \item \textbf{Heatmap} based on the normalized, cumulative attention measured to all parts of the visualization. 
    This revisualization recolors each face of the 3D geometries in the visualization based on the cumulative amount of attention the face has received, normalized between zero and the maximum amount of attention. 

    \item \textbf{Emphasis/de-emphasis} based on increasing and decreasing the color saturation of marks in the visualization. 
    The emphasis effect increases salience by recoloring the mark into a bright yellow, while the de-emphasis effect lowers visual salience by desaturating the mark through a filter. 
    Our implementation currently uses the short-term attention for this approach.

\end{itemize}

The 3D implementation follows the triggering strategies laid out in \cref{sec:impl-triggering}, and each of the revisualizations could be applied to each of the triggering strategies. 
However, we have currently only implemented emphasis/de-emphasis for {\scshape implicit} triggering.

\subsection{Implementation Notes}
\label{sec:3d-impl-notes}

\rev{Our 3D visualization implements bespoke visualizations in a WebXR application using the Three.js library.\footnote{\url{https://threejs.org}}
Attention is tracked through GPU color-picking on a per face basis; i.e., each triangle of every object's geometry is assigned a unique RGB color in the picking scene.
The red RGB colorspace is reserved for object IDs, while the green and blue RGB colorspaces are used for faces.
A benefit of using GPU-picking is that it is independent of the data geometry; this means we can even use custom 3D models.
With this approach we also do not need to be concerned with the layout of the 3D scene, and we can more easily get objects from an area compared to using a raycaster.}

The current implementation has a few limitations. 
Firstly, GPU picking is in itself an expensive operation. 
\rev{To improve performance, we only perform color-picking every 100 milliseconds, rather than every frame.}
Secondly, the use of colors limits the number, and complexities, of objects in the scene, and thus the size of the dataset to be visualized.
This has not been a problem for the datasets we have used.

\section{Evaluation Study}
\label{sec:study}

To validate our proposed solution to the research questions RQ1--RQ3 (\cref{sec:aav}), we carried out a qualitative evaluation involving human participants, presented in this section.
Our goal is to understand the design choices and deriving optimal strategies for visualizing attention as well as the associated triggering mechanisms. 

\subsection{Study Design Rationale}

We made the following decisions for our design study:

\begin{itemize}
    \item\textbf{No baseline.} We choose not to conduct a comparative or graphical perception experiment, partly because no suitable baseline comparison exists, and partly because we are interested in the richer guidance that a qualitative interview study provides.

    \item\textbf{Experienced participants.} We opted to recruit participants knowledgeable about data visualization to ensure that their attention would be representative of experienced visualization usage.

    \item\textbf{Convenience population.} Since our goal is to collect initial qualitative usage data of attention-aware visualizations, we feel that our specialized population (e.g., students in computer science and engineering) is of minor concern.

    \item\textbf{Between-participants on platform.} To keep sessions short, we split participants so that half did the study using the 2D data-agnostic AAV module, and half using the 3D data-aware one.

\end{itemize}

\begin{figure}[htb]
    \centering
    \subfloat[AAV2D condition.]{
        \includegraphics[width=0.48\columnwidth]{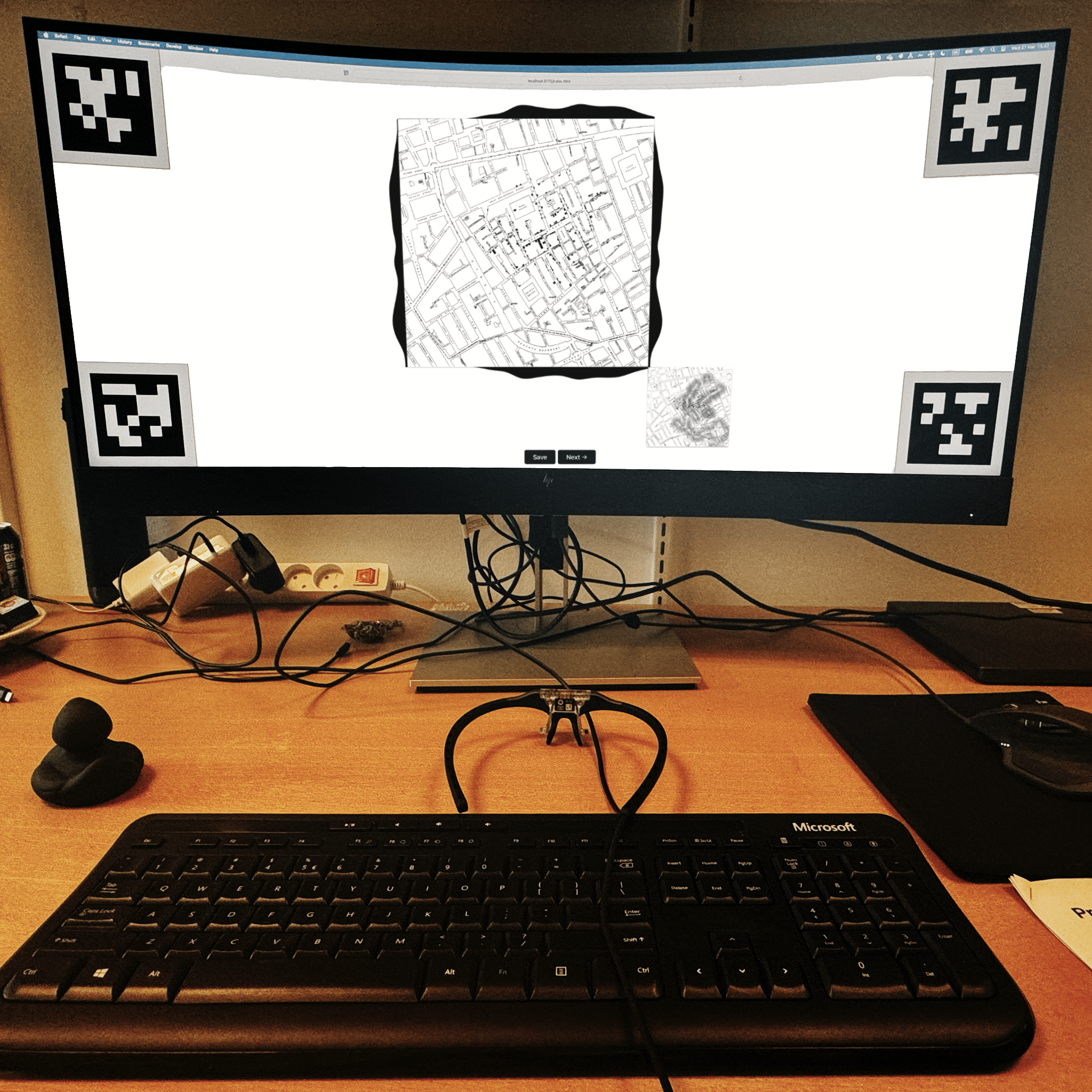}\label{fig:aav2d-interface}}
    \subfloat[AAV3D condition.]{
        \includegraphics[width=0.48\columnwidth]{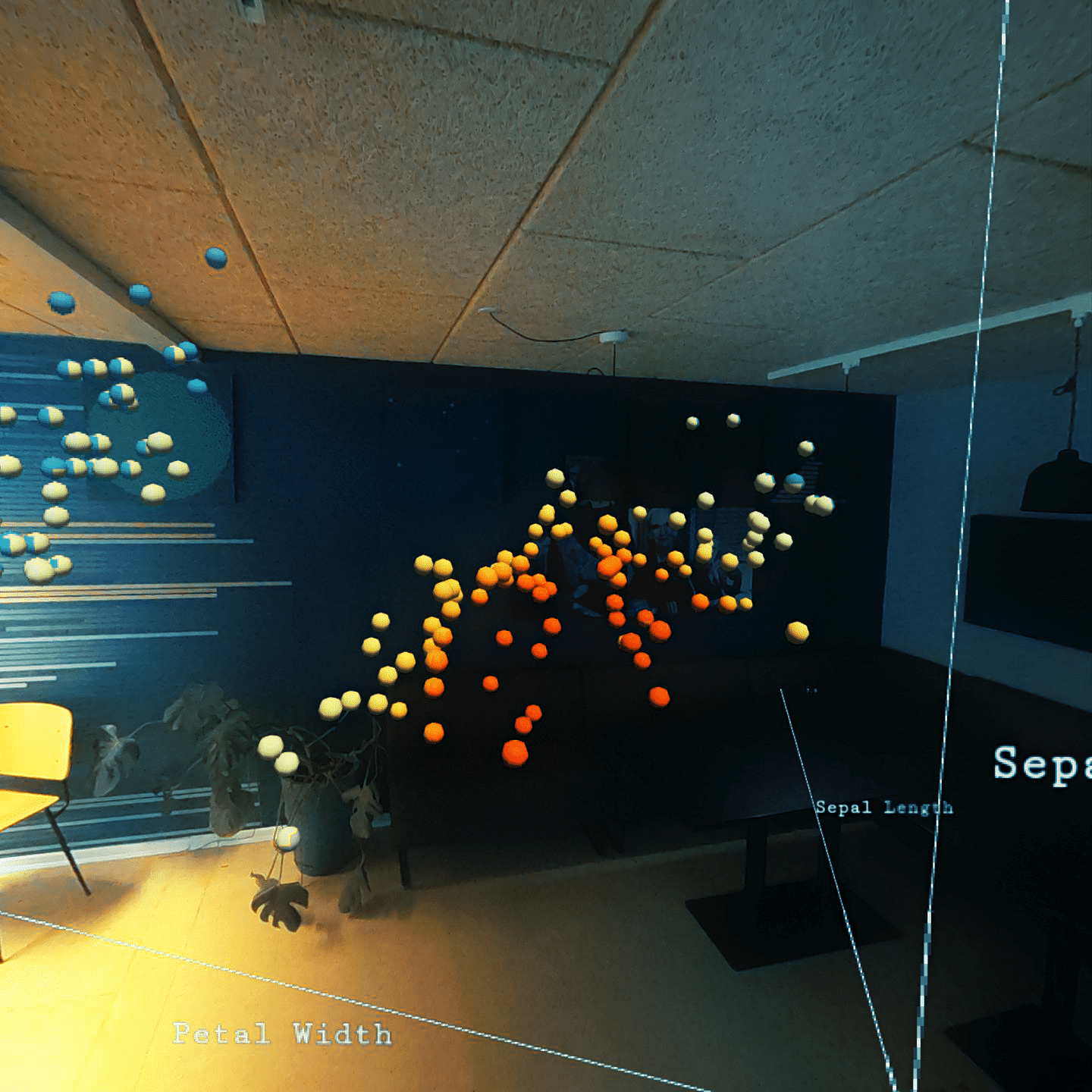}\label{fig:aav3d-interface}}
    \caption{\textbf{Experimental interface.}
    2D and 3D user study setups.}
    \label{fig:aav-interface}
\end{figure}

\subsection{Participants}

We recruited 12 participants (9 male, 3 female) from a convenience population (students at two different universities).
The participants were screened to be experienced computer users and knowledgeable about \add{reading and interpreting} data visualizations\add{ through a self assessment questionaire}.
Their age ranged from 24 to 52 (mean 30), all were enrolled in a computer science program, and all had normal or corrected-to-normal vision with no color vision deficiencies. 

\subsection{Apparatus}

For the 2D group, we used the data-agnostic implementation of attention-aware visualization in 2D (\cref{sec:2d-impl}).
We ran the experiment on an Apple Macbook Pro connected to a 34 inch ultra-wide monitor.
The software was displayed in a maximized web browser in a $3024\times1964$ pixel native resolution.
We used a Pupil Labs Neon embodied eyetracker connected to an Android smartphone, to process gaze data that was then fed into the AAV2D module on the computer.
\Cref{fig:aav2d-interface} shows an image of the experimental setup.

For the 3D group, we used the data-aware implementation of attention-aware visualization in 3D (\cref{sec:3d-impl}).
This study was conducted using a Meta Quest Pro head-mounted display in video passthrough mode.
\Cref{fig:aav3d-interface} shows the experimental setup.

\begin{figure*}[tbh]
    \centering
    \subfloat[\centering Charles Minard's \textit{Napoleon's March} (English Translation) (AAV2D).]{
        \includegraphics[height=3.5cm]{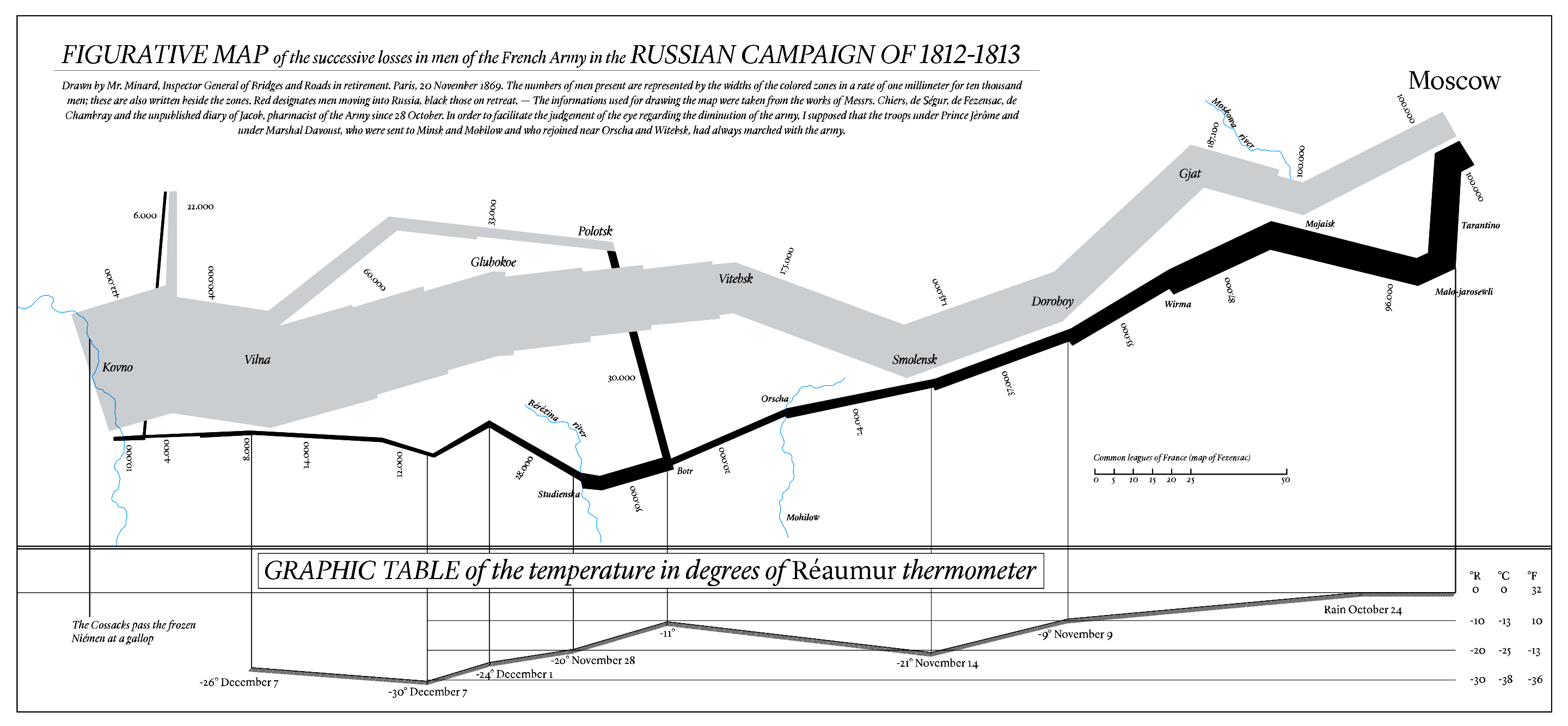}
        \label{fig:vis1}}
    \subfloat[\centering Florence Nightingale's \textit{Coxcomb Plots} (AAV2D).]{
        \includegraphics[height=3.5cm]{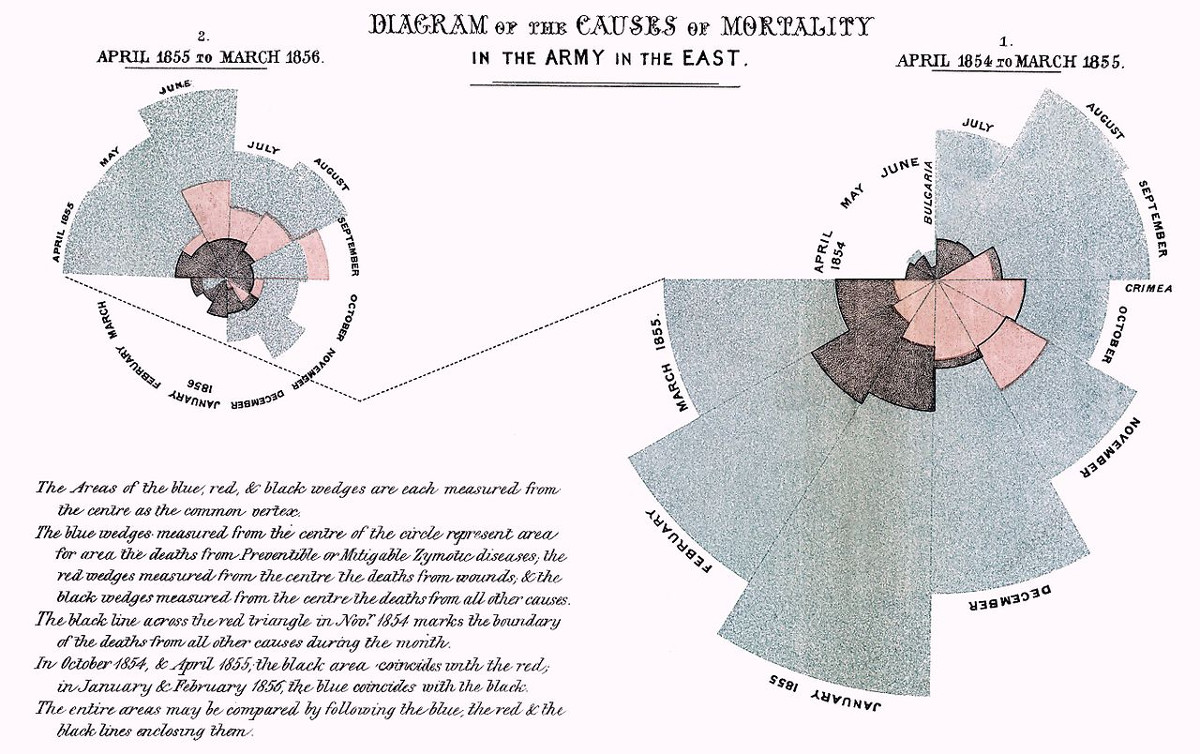}
        \label{fig:vis2}}
    \subfloat[\centering John Snow's \textit{Cholera Map} (AAV2D).]{
        \includegraphics[height=3.5cm]{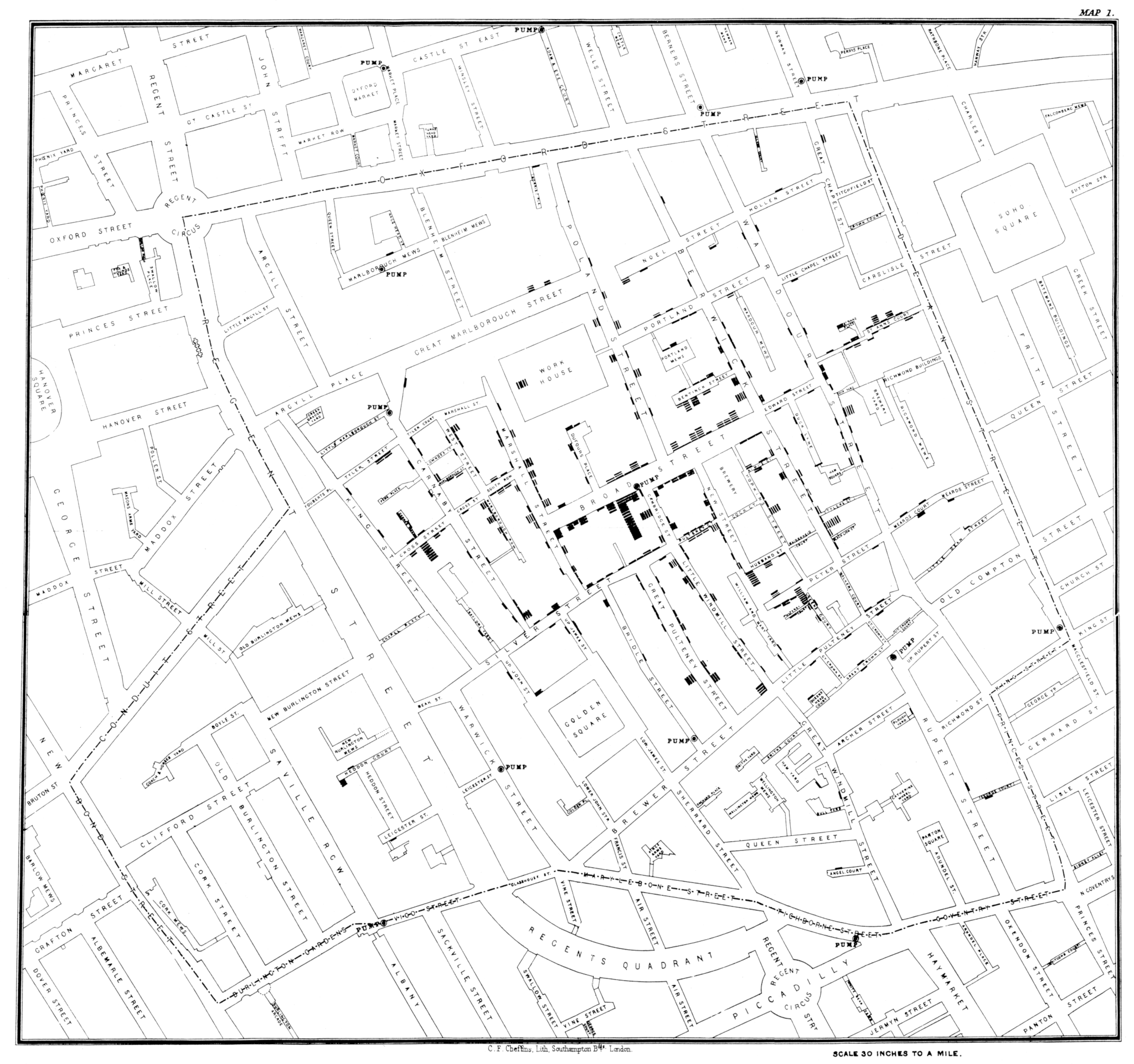}
        \label{fig:vis3}}\\
    \subfloat[\centering Fisher's \textit{Iris flower dataset}  as a 3D scatterplot (AAV3D).]{
        \includegraphics[width=.32\linewidth]{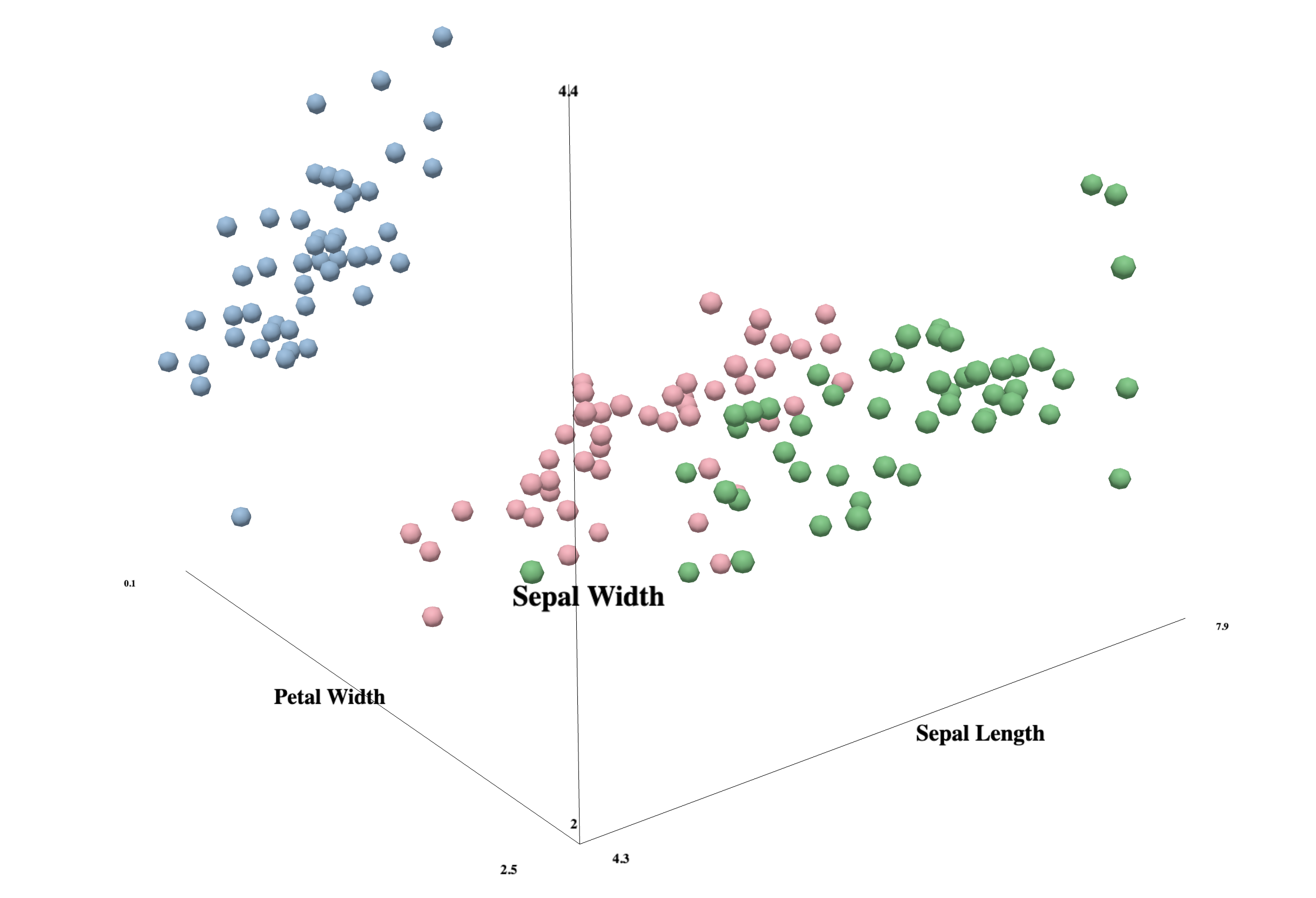}
        \label{fig:vis4}}
    \subfloat[\centering Quinlan's \textit{Auto MGP dataset}  as a 3D barchart (AAV3D)]{
        \includegraphics[width=.32\linewidth]{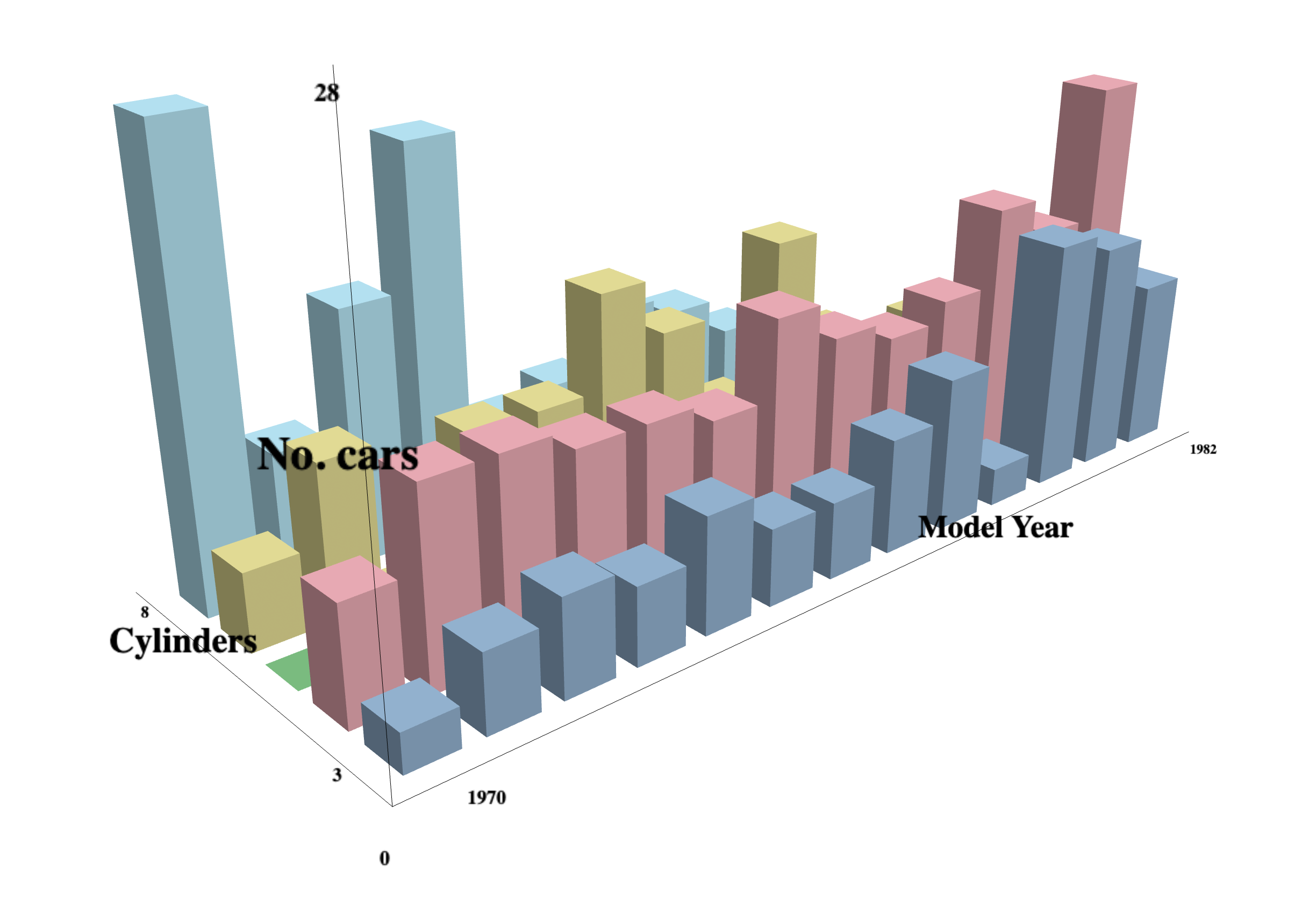}
        \label{fig:vis5}}
    \subfloat[\centering \textit{Mt.\ Bruno elevation} 3D surface (AAV3D).]{
        \includegraphics[width=.32\linewidth]{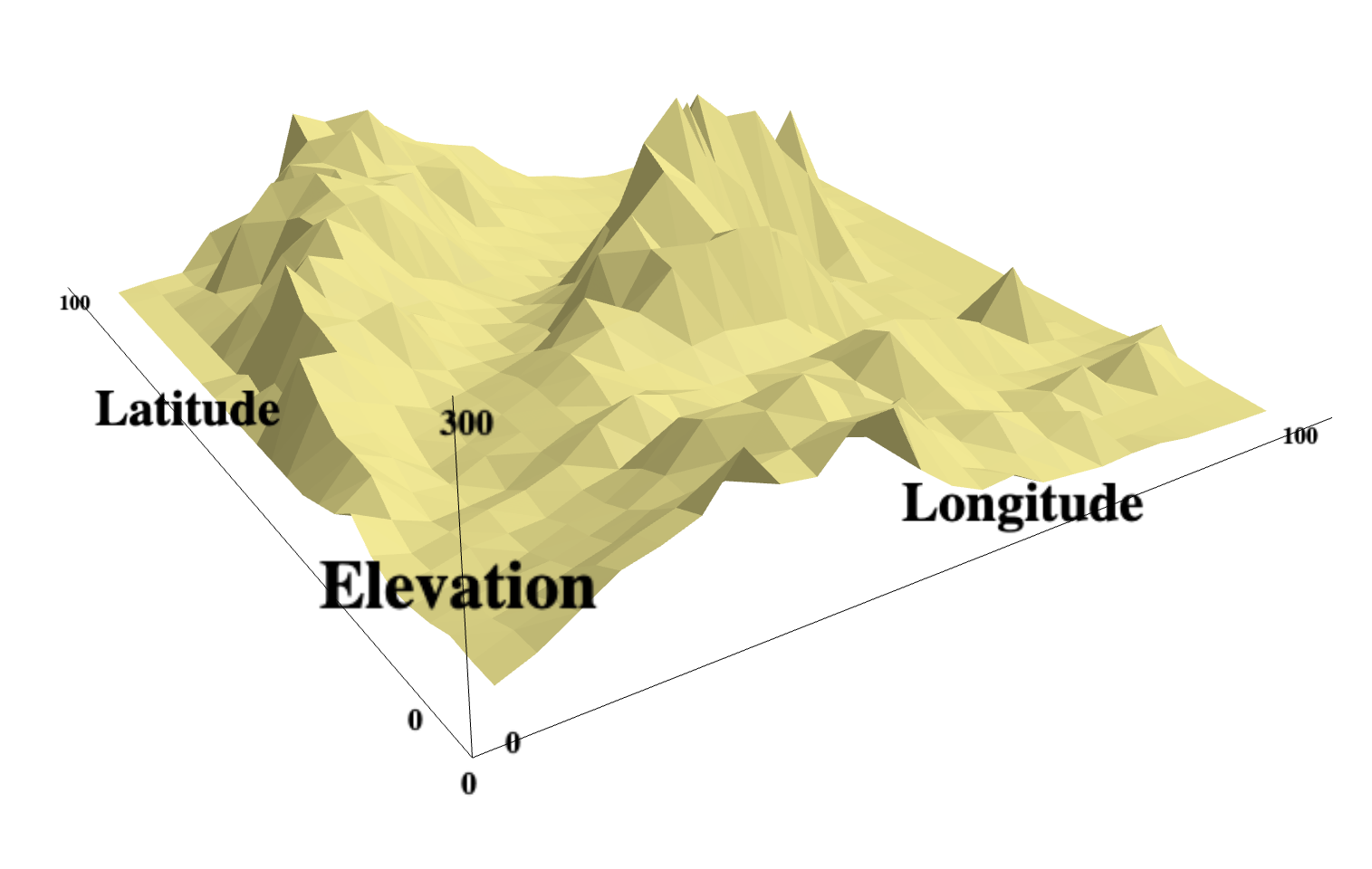}
        \label{fig:vis6}}
    \caption{\textbf{Study visualizations.}
    Each of these six visualizations (three for the 2D and three for the 3D platform) were associated with four tasks each (see supplemental material).
    All three of the 2D visualizations (top row) are in the public domain, and are slightly enhanced to improve readability.
    The 3D visualizations were our own designs.
    }
    \label{fig:study-vis}
\end{figure*}

\subsection{Procedure}

Participants first provided informed consent at the beginning of their evaluation session.
They filled out a demographic form and were given a brief introduction.
They were then asked to complete three task blocks in random order.
Each task block consisted of a visualization and a list of four open-ended questions.
\add{For the 2D tasks, participants engaged with 3 visualizations: Charles Minard’s Napoleon March (\cref{fig:vis1}), John Snow’s Cholera Map (\cref{fig:vis2}), and Florence Nightingale’s Rose Chart (\cref{fig:vis3}).
They were asked questions warranting exploration, such as identifying river crossings by Napoleon’s army, counting the number of pumps contaminated by cholera, and comparing mortality rates from diseases across different years.
For the 3D tasks, participants worked with a scatterplot of R.A.~Fisher's Iris data set (\cref{fig:vis4}), a barchart of Quinlan's Auto MPG dataset from UC Irvine (\cref{fig:vis5}), and an elevation map of Mount St.~Bruno (\cref{fig:vis6}).
They were asked to analyze car cylinder counts over time, identify the largest flower sepals, and compare elevations on the terrain map.
Detailed questions for these tasks can be found in the supplemental materials (see \cref{sec:supplemental_materials}).}
Participants were assigned a random combination of attention presentation and triggering mechanism.
\Cref{tab:conditions} gives the experimental conditions, three for 2D and 3D each.
Participants were given unlimited time to find answers to questions. 
Their performance was not timed and correctness was not measured.
Participants were encouraged to verbalize their thinking, but did not follow a strict think-aloud protocol. 

During each task block, the AAV software would continuously collect the user's gaze using the Neon eyetracker or using GPU picking based on the Meta Quest head direction.
The attention revisualization was displayed depending on the trigger mechanism randomly assigned to the participant: either manually or automatically.
Participants were shown how to trigger and use the attention display. 

After each block, participants were administered a brief System Usability Scale (SUS)~\cite{brooke1996sus} form and questionnaire on their experience and use of the attention-aware visualization.
After the study, they were administered a summary questionnaire and were interviewed on their experience.
Beyond these responses, we also captured attention metrics as well as usage statistics for each participant.

\begin{table}[htb]
    \renewcommand{\arraystretch}{1.2}
     \sffamily 
          \small
    \centering
    \caption{\textbf{Experimental conditions.}
        Six combinations of presentation and triggers used for the experiment; three per platform (2D and 3D).
    }
    \label{tab:conditions}
    \begin{tabular}{llll}
        \toprule
        \textbf{Condition} & \textbf{Platform} & \textbf{Trigger} & \textbf{Presentation}\\
        \hline
        2D-Always & AAV2D & Always-on/Explicit & Contour\\
        \rowcolor{SteelBlue!20}
        2D-Implicit & AAV2D & Implicit/Explicit & Mesh\\
        2D-Explicit & AAV2D & Explicit/Always-on  & Heatmap\\
        \rowcolor{SteelBlue!20}
        \hline
        3D-Always & AAV3D & Always-on & Heatmap\\
        3D-Implicit & AAV3D & Implicit & Saturate\\
        \rowcolor{SteelBlue!20}
        3D-Explicit & AAV3D & Explicit & Heatmap\\
        \hline
    \end{tabular}
\end{table}

\subsection{Results}
\label{sec:results}

Overall, all participants understood the visualizations and were able to complete the tasks.
Since we are not interested in completion time or task accuracy, we do not report such results here.
Instead, here we summarize the results of the qualitative aspects of our evaluation, including observations, the subjective ratings, and themes from the spoken and written participant feedback. 

\subsubsection{Observations}

All participants were able to complete all the tasks. 
We made no specific direct observations regarding task performance during either the 2D or 3D study.
We observed that the Pupil Labs Neon eye-tracker used for the 2D study was not completely accurate. For some participants it measured slightly below their actual gaze, and for others slightly above, which likely skewed the attention tracking. 
Although this slightly offset measured attention on the visualization potentially led to inaccurate revisualizations, we did not observe this being an issue for our study, given that the tasks were not dependent on revisualization accuracy, and the participants did not voice any accuracy concerns.

\subsubsection{Subjective Ratings}

\begin{figure}[htb]
    \centering
    \includegraphics[width=\linewidth, trim= 0 0 300 0,clip]{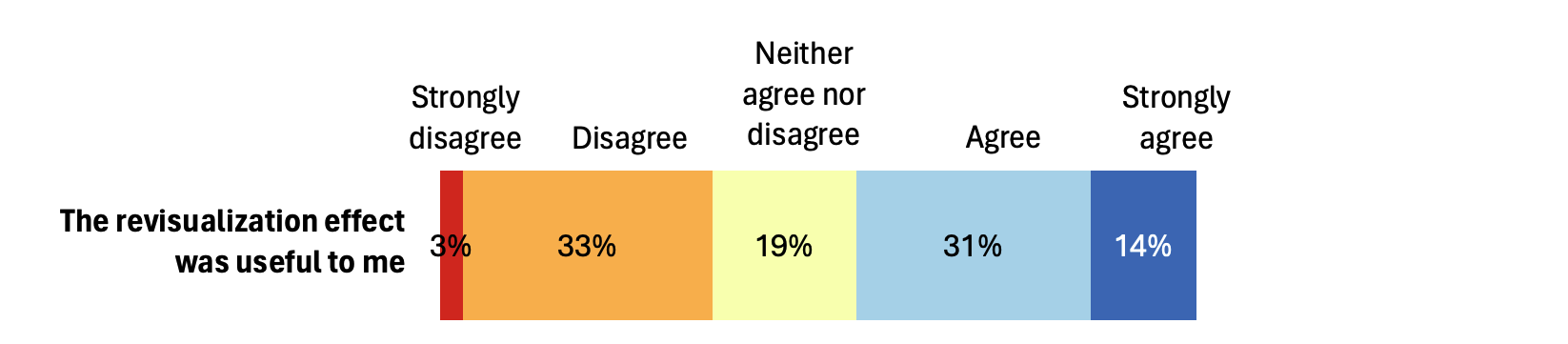}
    \caption{\textbf{Subjective usefulness.}
        The reported usefulness of the revisualizations was widely dispersed, with slightly more in favor of the revisualizations while a large portion of the participants were neutral. 
    }
    \label{fig:usefulness}
\end{figure}

The results for the SUS questionnaire were overall just above average~\cite{Bangor_SUS}, with the average score across participants of 70 (SD = 15). 
Per condition scores were all around 65 (below average SUS score), with one notable exception, the 3D visualization with \textsc{explicit} triggering, which scored 86 (excellent SUS score). 
When asked to rate the usefulness of the revisualizations, the responses varied significantly, with slightly more positive than negative responses. \cref{fig:usefulness} illustrates the reported usefulness of all the revisualizations. 
These dispersed responses aligns with the SUS questionnaire results. 

When asked to rank the triggering and revisualization techniques, participants using the 3D implementation had a clear preference for the \textsc{explicit} triggering condition and disliked the \textsc{always-on} heatmap. 
They preferred having control over when the revisualization was triggered, noting that the constant recoloring in the \textsc{always-on} condition was highly distracting, particularly because it overrode the colors of the visualization itself. 
This preference for \textsc{explicit} triggering is supported by the SUS score. 
The \textsc{implicit} triggering conditions were considered hard to understand, with one participant referring to it as ``mysterious.'' 
At the same time, other participants considered \textsc{implicit} triggering useful for certain tasks by helping to direct their attention.  

The participants using the 2D implementation did not have a clear preference for revisualization and triggering. 
While some participants liked the contour map for its precise marking of attention and disliked the fuzzy nature of the heatmap, other participants disliked how the contour map cluttered the minimap with lines and liked how the heatmap was a more subtle color overlay. 
Most participants did not like the Mesh condition. 
Reasons ranged from being harder to understand what was seen, to being annoyed that it made the minimap hard to read, since it darkened seen areas too much in their opinion, and was imprecise due to the bin size used. 

\subsubsection{Participant Comments and Feedback}

Our examination of participants' experiences with revisualization tools in 2D and 3D contexts not only revealed the significant impact of revisualizations on user engagement, attention, and analytical depth but also shed light on design preferences and the nuanced effects of these tools on user interaction with data visualizations.
We used thematic analysis to distill these insights into several key areas, incorporating participants' voices to ground our findings with additional context.

\paragraph{Guiding Attention and Informing Curiosity.}

Revisualizations were predominantly used by participants to refocus and redirect their attention, often illuminating areas within the visual space that had either been overlooked or required further exploration.
This utility of revisualizations bridged both 2D and 3D contexts, enhancing users' exploration thoroughness by guiding them towards both explored and unexplored areas (P2--P5 in 2D; P1, P3, P5 in 3D).
A participant (2D, P2) highlighted this dual functionality, noting, \textit{``It was more that it guided me to where I had been looking and which, of course, did tell me where I haven’t been looking.''}
Similarly, the stimulation of curiosity, especially when participants were seeking specific details or understanding the breadth of visualization techniques, was a recurring theme (P2, P5 in 2D; P2, P3 in 3D).
Some participants in 3D shared that they \textit{``treated it like a game''}(3D, P4) and \textit{``found (themselves) wanting to go inside the visualization and see it from inside,''} (3D, P2), underscoring the depth of engagement revisualizations can foster.

\paragraph{Subtlety, Clarity, and Control.}

Participants leaned towards a preference for revisualizations characterized by subtlety, clarity, and the provision of \textsc{explicit} control.
For instance, more than a couple of participants in 2D appreciated the subtlety of \textit{``shading (frame) around the screen''} (P1), with some stating that it \textit{``didn't disturb in situations where (they) choose to not use them''} (P3, P5).
Highly dynamic, vivid, or cluttered visualisztions were often met with disapproval due to their propensity to distract or even obscure crucial data points (P5 in 2D; P2, P6 in 3D).
For instance, more than a couple of participants in 2D remarked on the distracting nature of the Mesh implementation, stating how \textit{``the (Mesh) was very contrasting and changing... added too much clutter...''} (P5) and \textit{``occluded parts of the visualization''} (P2), even leading to some explicitly disliking it (P4).
Similar responses were seen from the 3D participants where they stated that \textit{``vibrant colors''} (P2) and \textit{``overriding or changing colors frequently''} seemed \textit{``harmful to experience''} (P6).
The preference for \textsc{explicit} control mechanisms over \textsc{always-on} features was also notably prevalent, as they were seen to reduce distractions and enhance user autonomy in data exploration (P1, P3 in 2D; P1, P3, P6 in 3D).
\textit{``Explicit is not distracting at all; I had control,''} a participant (3D, P3) reflected, highlighting the value of user-driven engagement with revisualization tools.

\paragraph{Subjective Preferences.} 

Participant feedback also pointed to the highly subjective nature of revisualization preferences and the critical role of task suitability in the effectiveness of these tools.
This was evident in the 2D context, as some participants gravitated towards specific revisualization techniques such as heatmaps or contour maps based on their orientation towards precision and their detail-oriented utility, others valued minimaps for providing a holistic perspective of the data (P2--P6 in 2D; P3, P5, P6 in 3D) 
\textit{``I did use the heatmap and the contour map more because I found them easier to understand how they can help me,''} noted a participant (2D, P3), exemplifying the varied preferences among users.
This diversity underscores the need for adaptable revisualization techniques that can cater to the preferences of individual users and specific analytical tasks.

\paragraph{Learning and Adaptation Over Time.}

Participants increasingly engaged with the revisualization techniques as the study progressed. 
Initially drawn to these tools by their novelty, participants experienced a rise in their comfort and efficiency in using them over time (P4 in 2D; P1--P3, P5 in 3D).
One participant in the 2D setting remarked, \textit{``At first, I was really paying attention because it was new... but after getting familiar, it feels routine and I hardly give it a second thought''} (P4 in 2D).
Similarly, in the 3D setting, another participant noted, \textit{``Indeed, as I became more accustomed to the effects, I found my attention more effectively directed,''} highlighting the dynamic process of their interaction with the tools (3D, P1).
This suggests that over time, users are likely to not just get used to the revisualization features but also incorporate them more fluidly into their analytical processes.

\section{Discussion}
\label{sec:disc}

We have validated our approach to attention-aware visualizations both through two separate implementations---one in 2D and one in 3D---as well as through a qualitative evaluation study.
Here we discuss what this work means for data visualization and context-aware designs within this field.
We also enumerate some of the limitations of our work.

\subsection{Explaining the Results}

Overall, our findings mostly confirmed our expectations: participants perceived attention awareness to be both novel and useful, and confidently incorporated these mechanisms into their analytical workflow.
As expected, however, the triggering and revisualization of attention is an important design challenge that requires careful attention (no pun).
On the other hand, there were some nuanced differences observed between the 2D and 3D implementations that are worth closer scrutiny.


The immersive quality of 3D environments accentuates the need for revisualization mechanisms, likely due to the increased potential of the immersion to overwhelm or distract.
\textit{``Quite like being able to see 3D data from a new perspective inside the headset, the difference of being able to walk around to see things from different angles rather than rotating,''} remarked a participant (3D, P3), highlighting the new perceptual possibilities made accessible through the addition of depth.
During its use, a participant commented, \textit{``I feel like my attention is being guided around the visualizations; I have to actually move around to paint the scene, encouraging exploration''} (3D, P6), indicating that revisualization mechanisms can become essential for navigating and understanding data in 3D spaces by serving as a counterbalance to the potential for sensory overload and confusion.
In contrast, in 2D environments, where depth is absent and the entire visualization is fully visible at all times, the need for revisualization appears less critical.

These insights suggest that the design of AAVs might benefit from a nuanced understanding of how users interact with and perceive visualizations differently in 2D vs.\ 3D.
While the immersive complexity of 3D environments could amplify the need for user control and engaging interaction mechanisms, 2D spaces might call for simpler and more direct approaches.
Anticipating and adapting to these differences could be key to unlocking the full potential of AAVs, ensuring that they enhance rather than complicate the user experience.


\subsection{Design Implications}

The introduction of attention-aware visualizations (AAVs) marks a shift towards creating more interactive and personalized data exploration experiences.
By leveraging user attention data to dynamically adjust visualizations, we think that AAVs can make visual data exploration more tailored to individual users' needs.
However, our study highlights a critical balance that must be struck to harness the full potential of AAVs effectively.
Continuous active monitoring and adjustment based on user gaze can be disruptive, creating a feedback loop that might skew the user's natural data exploration process.
Conversely, relying solely on user-initiated triggers for attention-aware adjustments risks underutilization due to forgetfulness or the extra cognitive load of remembering to activate the feature.
This underscores the need for AAV systems to intelligently discern when to provide guidance without overwhelming the user.\footnote{After all, no one wants a visualization version of Clippy~\faPaperclip.} 
Such a system would ideally analyze patterns of user engagement to anticipate the need for attention redirection, offering a seamless blend of automated and user-controlled interactions.

\add{Even though we have only implemented and evaluated two of the settings that AAV could be useful in, we feel that AAV has potential in other settings as well, such as attention-aware 2D visualizations in immersive environments, or 3D visualizations viewed in a non-immersive environment, i.e., on a screen. 
Non-immersive 3D visualizations are arguably more common than their immersive counterparts, which is why our approach in AAV3D is designed to also work for such settings. 
We chose to focus on an immersive environment in this paper because we believe that AAVs can provide especially great value for such settings. 
When data is distributed all around the user, the cognitive load of keeping track of the data is high. 
In such scenarios, AAVs could be used to reduce the cognitive load, by pointing to unseen areas, or by reminding the user if they have already analyzed certain areas of the data. 
While a data-aware 2D implementation will not cause any difficulty, there are issues associated with a data-agnostic 3D implementation using the current approach. Our 2D implementation captures attention as a grid overlay. If the user is allowed to rotate view of the visualization and thereby the position of the marks on the screen, the stored attention will no longer correspond to the correct marks, since they can be at another cell in the overlay. This is a general issue if the user is able to change the representation. 
How to address this issue is left to future work.}

Looking to the bigger picture, the integration of AAVs into visualization design may help improve user engagement with data.
One potential direction involves the development of context-aware AAVs that adapt not just to where users look but also to the context of their interaction---what they are looking for and their task at hand.
By understanding the user's intent and the context of their exploration, AAVs could provide more meaningful adjustments, such as highlighting relevant data points or suggesting alternative visualizations that might offer deeper insights into the data.
This level of adaptation requires advancements in machine learning and artificial intelligence to interpret complex patterns of attention and interaction, paving the way for visualizations that are not only reactive but also proactive in facilitating data discovery.

Finally, the step to collaborative and educational uses for attention-aware visualization is not far.
In situations where multiple users interact with the same visualization, AAVs could dynamically adjust to accommodate the focus and needs of different users, enhancing collaborative analysis and decision-making.
In educational contexts, AAVs could tailor the presentation of information to the learner's pace and points of interest, providing a more engaging and effective learning experience.

\subsection{Limitations}

As discussed in our design framework (\cref{sec:aav}), one limitation of our approach is the potential for reinforcing behavior, where revisualizing mechanism based on user gaze may inadvertently influence the user's subsequent attention patterns.
Such feedback could lead to a self-reinforcing cycle, where the visualization continuously emphasizes areas previously focused on by the user, potentially overshadowing other valuable but less attended data points.
Understanding and mitigating the impact of such reinforcing behavior on the user's exploration and discovery process remains a challenge for future iterations of AAVs.

Another limitation concerns the choice of triggering to avoid disrupting the user experience.
Finding the right balance between providing useful feedback and avoiding overwhelming the user with frequent or intrusive adjustments to the visualization is critical.
This requires careful consideration of the timing, frequency, and extent of revisualizations to ensure they enhance rather than detract from the data exploration process.
We leave such fine-tuning and evaluation to future work.

While eye tracking technology is common in the current generation of AR/VR headsets, it is still not commonplace in typical computing setups, thus limiting the adoption of AAVs.
This restricts the immediate applicability of our proposed methods and limits the user base that can benefit from them.
We speculate that the prevalence of gaze interaction in, e.g., the Apple Vision Pro will lead to such technologies also becoming popular for regular computers and mobile devices.

The informal nature of our evaluation represents a methodological limitation.
The \rev{evaluation}'s qualitative focus, lack of quantitative metrics, and absence of a baseline comparison restrict our ability to draw definitive conclusions about the efficacy of the framework.
Further research involving more rigorous, quantitative evaluations and comparisons with existing approaches is necessary to fully understand the benefits and limitations of attention-aware visualization.

\add{While we envision numerous futures for AAV, realizing them means developing algorithms capable of managing the delicate balance between being informative and non-disruptive.
These algorithms are needed to ensure that AAVs enrich the user experience by making data visualizations not just tools for viewing but also guiding exploration.}

\section{Conclusion and Future Work}
\label{sec:conclusion}

We have presented attention-aware visualizations (AAVs), a context-aware data visualization paradigm that leverages user attention to dynamically adapt visual content.
By tracking and analyzing where and how users allocate their attention over time, AAV not only offers a more personalized data exploration experience, but also detects situations where the user is not seeing everything in a given visualization.
Our approach has been operationalized through two distinct implementations: a 2D data-agnostic method suitable for web-based visualizations and a 3D data-aware technique utilizing the stencil buffer for immersive analytics environments.
Both methods underscore the potential of integrating attention metrics to enhance the interpretability and user engagement of visualizations.
Accordingly, our evaluation of these implementations has provided insights into the effectiveness of visual feedback and attention-triggering mechanisms, highlighting the utility of AAVs in facilitating a deeper understanding of complex datasets.

We feel that this work points to future context-aware visualization techniques, where the focus is on developing visualization systems that respond to the needs of their users.
Exploring the integration of AAVs with other sensory input methods, such as auditory feedback or haptic interfaces, could offer a richer, more multisensory data interaction experience.
Additionally, the application of machine learning techniques to predict user attention shifts could refine the responsiveness of AAVs, making visualizations even more intuitive and user-centric.
Another promising direction involves investigating the scalability of AAVs in handling larger, more complex datasets and their applicability across diverse fields such as medical imaging, financial analysis, and social network visualization.
Finally, conducting extensive user studies across different domains could shed light on the universal applicability of AAVs and tailor them more closely to user needs and preferences.


\section*{\add{Supplemental Materials}}
\label{sec:supplemental_materials}

\add{All supplemental materials are available on OSF at \url{https://osf.io/8mfhp/}, released under a CC BY 4.0 license.
In particular, they include the following:
(1) Analysis files containing the final results of the quantitative System Usability Rating (SUS) analysis for both the 2D and 3D studies, along with a consolidated and summarized qualitative analysis with corresponding quotes from participants;
(2) Data files containing the collected data from the participants of our evaluation. 
These files include experimental data in the form of data snapshots (only for 2D) and participant responses consolidated from block-level and study-level surveys, supported by notes and quotes from post-study interviews;
(3) Materials used during the evaluation, including demographic forms and questionnaires; and 
(4) Zipped versions of the implementations of AAVs.}


\section*{\add{Figure Credits}}
\label{sec:figure_credits}

\add{Unless specified below, all figures are credited to the authors.}

\Cref{fig:vis1} \add{image credit: 
Modern redraw in English by Inigo Lopez Vazquez. May 3, 2015, CC BY-SA 4.0. Original by Charles Minard. November 20, 1869, public domain.}

\Cref{fig:vis2} \add{image credit: 
Florence Nightingale. 1858, public domain.}

\Cref{fig:vis3} \add{image credit: 
John Snow, 1854, public domain.}


\acknowledgments{%
    This work was supported partly by Villum Investigator grant VL-54492 by Villum Fonden.
    Any opinions, findings, and conclusions expressed in this material are those of the authors and do not necessarily reflect the views of the funding agency.
}

\bibliographystyle{abbrv-doi-hyperref}
\bibliography{aav}

\end{document}